\documentclass[review,12pt]{elsarticle}

\usepackage[margin=1.55cm]{geometry}
\usepackage{amssymb}
\usepackage{amsmath}
\usepackage[utf8]{inputenc}
\usepackage{enumerate}
\usepackage[table]{xcolor}
\usepackage{mathrsfs}
\usepackage{booktabs}
\usepackage{enumitem}
\usepackage{color, colortbl}
\usepackage{rotating}
\usepackage{booktabs}
\usepackage{multirow}
\usepackage{multicol}
\usepackage{mathtools}
\usepackage{verbatim}
\usepackage{amssymb}
\usepackage{setspace}
\usepackage{footnote}
 \usepackage{amsthm}
\usepackage{multirow}
\usepackage{algorithmicx}
\usepackage{tikz}
\usepackage{pgfplots}
\usetikzlibrary{calc}
\usetikzlibrary{intersections}
\usepackage{caption}
\usepackage{subcaption}
\usetikzlibrary{shapes.geometric}
\usetikzlibrary{shapes.arrows}
\usetikzlibrary{chains, shapes.misc}
\usepackage{array}
\usepackage{footnote}
\usepackage{longtable}
\usepackage{pdflscape}
\usepackage{tablefootnote}
\usepackage{subcaption}
\usepackage{eucal}
\usepackage{mathtools}
\usepackage{enumerate}
\usepackage{arydshln}
\usepackage{enumerate}
\usepackage{rotating}
\usepackage{caption}

\usepackage[T1]{fontenc}
\usepackage[utf8]{inputenc}

\usepackage{apalike}

\usepackage[linesnumbered,ruled,vlined]{algorithm2e}

\usepackage[flushleft]{threeparttable} 
\usepackage{booktabs,caption}

\usepackage{mathtools}

\usepackage{algpseudocode}
\usepackage{tabu}
\usepackage{tabularx}
\usepackage{xcolor, soul}
\sethlcolor{yellow}

\usepackage{lscape}
\usepackage{adjustbox}
\usepackage{xcolor}
\definecolor{darkgreen}{rgb}{0.0, 0.7, 0.0}
\definecolor{orange}{rgb}{1.0, 0.5, 0.0}
\definecolor{darkred}{rgb}{0.95, 0.0, 0.0}

\usepackage{newtxtext,newtxmath} 
\usepackage{pifont} 
\usepackage{natbib}

\usepackage{bm}      

\usepackage[normalem]{ulem} 

\usepackage{setspace}
\usepackage[colorlinks,citecolor=red,urlcolor=blue,bookmarks=false,hypertexnames=true]{hyperref}
\newcolumntype{L}[1]{>{\raggedright\let\newline\\\arraybackslash\hspace{0pt}}m{#1}}
\newcolumntype{C}[1]{>{\centering\let\newline\\\arraybackslash\hspace{0pt}}m{#1}}
\newcolumntype{R}[1]{>{\raggedleft}m{#1}}
\usepackage{hyperref}

\pgfplotsset{compat=1.18}
\begin{document}
\begin{frontmatter}

\title{Adaptive alert prioritisation in security operations centres via learning to defer with human feedback}

\author[a]{Fatemeh Jalalvand\corref{cor1}}
\ead{Fatemeh.Jalalvand@data61.csiro.au}

\author[a]{Mohan Baruwal Chhetri}
\ead{Mohan.Baruwalchhetri@data61.csiro.au}

\author[a]{Surya Nepal}
\ead{Surya.Nepal@data61.csiro.au}

\author[a]{C\'ecile Paris}
\ead{Cecile.Paris@data61.csiro.au}

\address[a]{CSIRO’s Data61, Australia}

\cortext[cor1]{Corresponding author}

\journal{}
	\begin{abstract}
		\small
        Alert prioritisation (AP) is crucial for security operations centres (SOCs) to manage the overwhelming volume of alerts and ensure timely detection and response to genuine threats, while minimising alert fatigue. Although predictive AI can process large alert volumes and identify known patterns, it struggles with novel and evolving scenarios that demand contextual understanding and nuanced judgement. A promising solution is Human-AI teaming (HAT), which combines human expertise with AI’s computational capabilities. Learning to Defer (L2D) operationalises HAT by enabling AI to “defer” uncertain or unfamiliar cases to human experts. However, traditional L2D models rely on static deferral policies that do not evolve with experience, limiting their ability to learn from human feedback and adapt over time. To overcome this, we introduce \textbf{Learning to Defer with Human Feedback} (L2DHF), an adaptive deferral framework that leverages Deep Reinforcement Learning from Human Feedback (DRLHF) to optimise deferral decisions. By dynamically incorporating human feedback, L2DHF continuously improves AP accuracy and reduces unnecessary deferrals, enhancing SOC effectiveness and easing analyst workload. Experiments on two widely used benchmark datasets, UNSW-NB15 and CICIDS2017, demonstrate that L2DHF significantly outperforms baseline models. Notably, it achieves 13-16\% higher AP accuracy for critical alerts on UNSW-NB15 and 60-67\% on CICIDS2017. It also reduces misprioritisations, for example, by 98\% for high-category alerts on CICIDS2017. Moreover, L2DHF decreases deferrals, for example, by 37\% on UNSW-NB15, directly reducing analyst workload. These gains are achieved with efficient execution, underscoring L2DHF’s practicality for real-world SOC deployment.

	\end{abstract}
\end{frontmatter}

\section{Introduction} \label{sec:Intro}

Effective alert prioritisation (AP) is crucial in security operations centres (SOCs) to help analysts identify and respond to critical alerts amid the overwhelming volume of daily notifications. Organisations often receive in excess of 10,000 alerts daily \citep{van2022deepcase,FireE}, making it easy for genuine threats to be overlooked among less important ones. The sheer volume of alerts, many of which are false positives, contributes significantly to analyst fatigue \citep{baruwaltowards,jalalvand2024alert}. When AP systems fail to reliably distinguish false positives from genuine attacks, analysts are inundated with low-priority or irrelevant alerts~\citep{Alahmadi2022-wb}. Worse, misprioritisinging critical alerts as medium or low priority constitutes a serious security hazard by allowing genuine threats to go undetected, while also increasing the analysts' cognitive load when such alerts must be re-investigated at a later stage. Together, these factors reduce the analysts' ability to focus on and respond effectively to genuine security incidents~\citep{baruwaltowards}.



To address these challenges, artificial intelligence (AI) and machine learning (ML)-enabled tools are increasingly adopted to automate AP tasks. However, despite their growing use, studies such as TrendMicro’s Global Study~\citep{SOC_TrendMicro_Survey} and SANS Institute’s SOC Survey~\citep{SANS2023} highlight that SOCs continue to face significant challenges in managing cyber threats. The dynamic nature of the threat landscape and the inherent uncertainty of security situations makes fully automated AP problematic. Even with recent advances in AI, ML-based systems remain brittle when confronted with novel or unexpected scenarios not captured in their training data~\citep{tariq2025a2c}. A key limitation is their inability to recognise when they are likely to fail~\citep{zhang2023survey}, often resulting in incorrect predictions with high confidence~\citep{tariq2025a2c}.

Such failures can cause the system to overlook critical alerts or mistakenly prioritise benign ones, thereby increasing the possibility of undetected attacks. According to MITRE~\citep{knerler202211}, while automated AP helps SOCs manage the overwhelming volume of raw alerts, human analysts remain indispensable for interpreting and contextualising these alerts. As noted, "automation assists, but does not fully replace, the judgement of advanced human analysts"~\citep{knerler202211}. 
One approach to incorporating human input into AP involves periodic retraining of AI models with analyst feedback~\citep{kim2022active,hore2022towards}. However, this approach lacks real-time adaptability, introduces operational inefficiencies due to repeated training cycles, and results in delayed integration of human expertise, limiting its effectiveness in dynamic threat environments.

Human-AI teaming (HAT) has the potential to improve AP by leveraging the unique strengths of human analysts and AI systems \citep{jalalvand2024alert}. HAT enables collaboration between human expertise and machine intelligence, producing outcomes neither could achieve alone \citep{baruwaltowards,cleland2022extending, schleiger2024collaborative}. AI excels at processing large volumes of alert data, detecting anomalies, uncovering hidden patterns, and prioritising alerts at scale. In contrast, human analysts contribute contextual understanding, validate AI-generated priorities, and provide valuable feedback to refine the system \citep{jalalvand2024alert}. This synergy  improves overall performance and reduces  misprioritisations \citep{jalalvand2024alert,tariq2025alert}. Furthermore, by minimising irrelevant alerts, HAT can help reduce alert fatigue, a persistent challenge in SOC environments \citep{baruwaltowards, tariq2025alert}.

Learning to Defer (L2D) \citep{madras2018predict} is a paradigm that operationalises HAT by allowing AI systems to defer decision-making in uncertain or complex situations to human experts. L2D offers a promising solution for AP by enabling AI to handle routine alerts autonomously while deferring uncertain or novel cases to human analysts for contextual judgements. In this setup, the AI first classifies\footnote{For the purpose of this study, we use prioritisation, categorisation, and classification interchangeably to refer to the process of assigning importance or priority to alerts.} incoming alerts, deferring those beyond its confidence threshold for human review. Human analysts then conduct further investigations, identify novel threats, and refine prioritisation decisions. L2D consists of two core components: a \textit{predictive model} that classifies alerts, and a \textit{deferral model} that determines whether to rely on the AI's output or defer to human judgement. By striking this balance, L2D enhances AP accuracy while reducing analyst involvement. However, current L2D implementations are limited by their static deferral mechanisms. Once trained, the deferral model lacks the ability to evolve based on ongoing human interactions, missing crucial opportunities for continuous learning and improvement. Moreover, this inflexibility can lead to repeated deferrals of similar alerts, increasing analyst workload and hindering system performance. Incorporating real-time feedback mechanisms for continuous adaptation is essential to optimise AP performance in dynamic cybersecurity environments \citep{tariq2025alert, jalalvand2024alert}.

To address this limitation of L2D, we propose \textbf{Learning to Defer with Human Feedback (L2DHF)}, illustrated in Figure \ref{PAI-DRLHF}. L2DHF enhances the traditional L2D approach by replacing the static deferral model with an adaptive deferral model, implemented as a Deep Reinforcement Learning (DRL) agent trained through human feedback, following the Deep Reinforcement Learning from Human Feedback (DRLHF) framework. This enhancement enables the deferral strategy to evolve over time based on real-time analyst feedback, supporting continuous learning and improved AP performance. Initially, the predictive AI assigns priorities to the incoming alerts, after which the DRL agent decides whether to accept the predictive AI's prioritisation or defer uncertain alerts to human analysts. Analysts apply their domain knowledge and contextual insights to validate and adjust priorities, providing feedback to the DRL agent. This feedback loop strengthens the DRL agent’s learning and improves its deferral decisions, thereby increasing AP accuracy, reducing misprioritisation between severity threat levels, including false positives and false negatives, and easing analyst workload. Additionally, the Analyst-Validated Alert Repository (AVAR) stores analyst-validated alerts, enabling the system to recognise and filter out previously seen and validated alerts, thus streamlining the prioritisation process.

\subsection{Contributions} \label{sec:Conts}
The main contributions of this study are:

\begin{enumerate}[label=\roman*.]
    \item \textbf{The L2DHF Framework:} We propose L2DHF, a novel extension of the L2D paradigm that integrates a dynamic and adaptive deferral mechanism implemented via a DRL agent within a DRLHF architecture. This enables an effective real-time HAT, allowing the DRL agent to continuously refine its deferral policy based on analyst feedback, thereby improving  decision-making over time.
    \item \textbf{Application of L2DHF to AP:} We apply L2DHF to the core SOC task of AP. L2DHF facilitates real-time, dynamic HAT for AP, improving AP accuracy, reducing analyst workload, enhancing SOC efficiency and effectiveness, and addressing key operational challenges.
    \item \textbf{Empirical evaluation:} We evaluate L2DHF using two widely adopted network intrusion datasets, UNSW-NB15~\citep{moustafa2017big,moustafa2015unsw} and CICIDS2017~\citep{sharafaldin2018toward}. Results demonstrate that L2DHF significantly outperforms baseline approaches. For instance, L2DHF improves AP accuracy for critical alerts by 13-16\% on UNSW-NB15 and 60-67\% on CICIDS2017. It also substantially reduces  misprioritisations, including a 100\% reduction in  misprioritisation of critical alerts, a 98\% reduction in  misprioritisation of high-category alerts, and a 52\% reduction in false positives on CICIDS2017. Furthermore, L2DHF reduces analyst workload, as evidenced by 37\% reduction in deferred alerts on UNSW-NB15 compared to baseline approaches.
\end{enumerate}

\subsection{Structure} \label{sec:struc}

The rest of this article is structured as follows. Section \ref{sec:relatedW} provides an overview of related work. Section \ref{sec:method} introduces the proposed L2DHF method, and Section \ref{sec:implement} discusses its implementation. Section \ref{sec: experimental setup} describes the experimental setup.
Section \ref{sec: results} provides the experimental results, followed by a presentation of threats to validity in Section \ref{sec: limit}. Finally, Section \ref{sec:conc} concludes the paper.

\begin{figure*}[!t]
  \centering
  \includegraphics[width=0.8\linewidth]{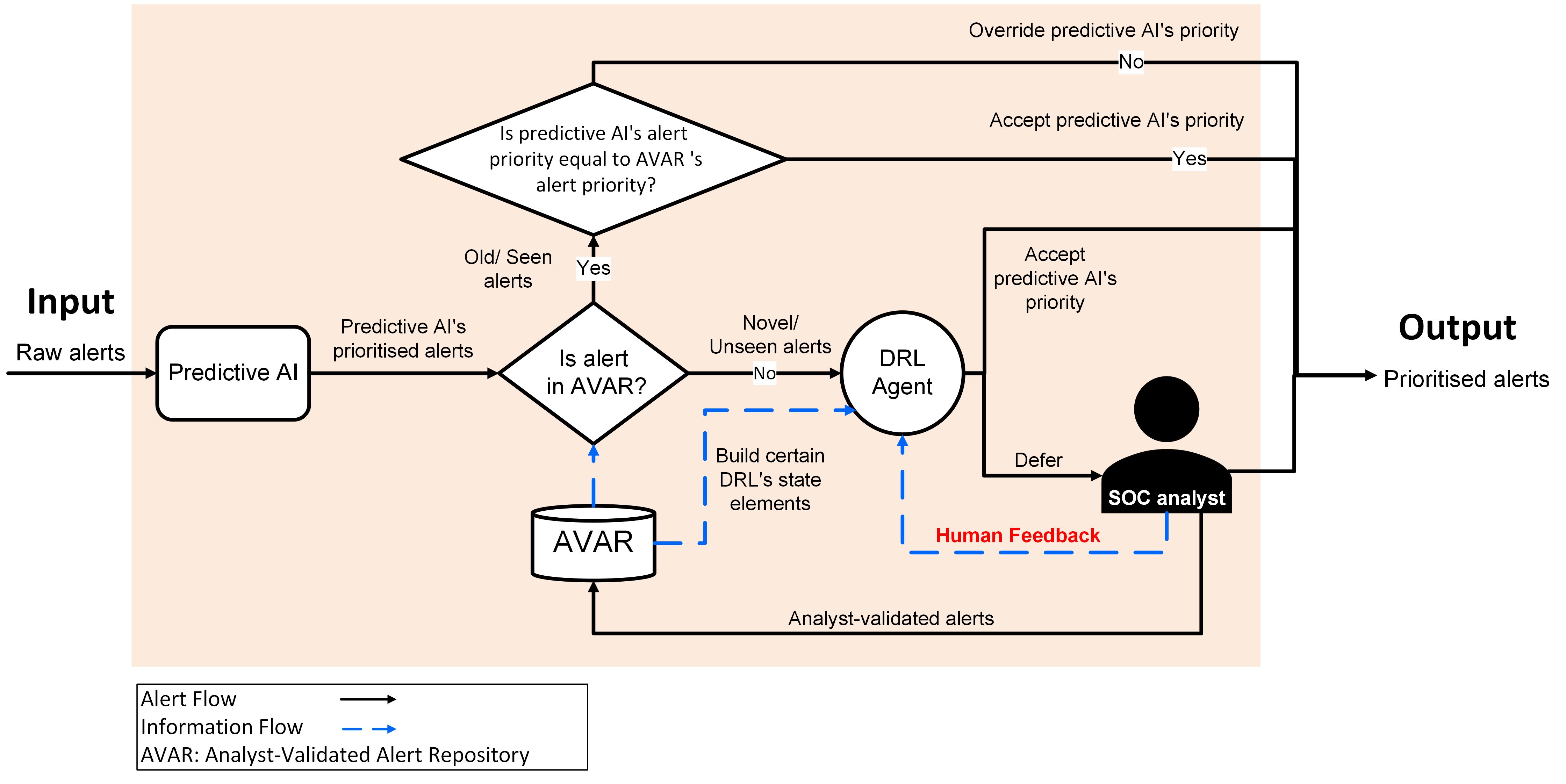}
  \caption{Learning to Defer with Human Feedback (L2DHF) framework.}
  \label{PAI-DRLHF}
\end{figure*}

\section{Related work and background} \label{sec:relatedW}
This section highlights the most relevant recent works on ML-based AI solutions as well as HAT for AP, aiming to identify research gaps and emphasise our contributions. It is not intended as an exhaustive review. For a comprehensive review on AP approaches in SOCs, including those addressing HAT, refer to \citep{jalalvand2024alert}.

\subsection{ML-based AI solutions} \label{AI for AP}

Extensive research has focused on automating AP using ML-based AI solutions, particularly supervised and unsupervised learning~\citep{ aminanto2020threat, jeamaon2022cybersecurity,ongun2021portfiler, feijoo2023cybersecurity,alperin2019risk}.
For example, \cite{ongun2021portfiler} developed an automated AP method by employing ensembles of various unsupervised ML techniques, such as Isolation Forest and unsupervised deep learning to generate anomaly scores for alert ranking. \cite{alperin2019risk} proposed a hybrid approach that integrates natural language processing (NLP) with supervised ML models such as random forests. In this approach, NLP techniques are used to extract key attributes from alert data, which are then fed into the supervised ML models to assign vulnerability scores used to prioritise alerts. 

Beyond supervised and unsupervised learning, several studies have investigated reinforcement learning (RL) approaches for automated AP \citep{chavali2022sac,hore2023deep,huang2022radams,tong2020finding}. For example,  \cite{hore2023deep} applied DRL to allocate resources required for mitigating vulnerabilities and subsequently employed an integer programming model to prioritise vulnerabilities based on the resource allocation determined by DRL. Similarly, \cite{tong2020finding} used game theory to model the defender–attacker interaction for AP and applied an adversarial RL framework to capture the computational complexity involved in solving the game.

The majority of SOC platforms integrate some form of  ML-based AI technologies to enhance SOC operations. In particular, AP is typically fully automated in these systems, with little to no human analysts involvement in the prioritisation process. Examples of state-of-the-art solutions that incorporate AI systems to automate AP as part of their service offerings include Trend Vision One \citep{TrendVisionOne}, Palo Alto Networks Cortex XSOAR \citep{Cortex_XSOAR}, and Splunk SOAR \citep{Splunk}. For instance, Palo Alto Networks Cortex XSOAR \citep{Cortex_XSOAR} leverages ML capabilities for automated AP. It uses classifiers to categorise alerts and assign corresponding priorities. These classifiers are trained using analyst-provided data, ensuring that the system meets the specific, customised requirements.

\subsection{Human-AI teaming paradigms} \label{HAT for AP}
HAT for AP is still in its early stages. A detailed examination of AP methods in SOCs shows that methods integrating HAT for AP are still largely unexplored \citep{jalalvand2024alert}. In this work, we frame our discussion of HAT for AP around three aspects: periodic AI retraining, L2D, and reinforcement learning from human feedback (RLHF), based on our critical review of the landscape.

\subsubsection{Periodic model retraining} \label{HAT-re-training}
Few research efforts have explored AI retraining to enable HAT for AP \citep[see, e.g., ][]{hore2022towards,gelman2023escalated,kim2022active,liu2022context2vector, hossain2020combating}. For instance,  \cite{gelman2023escalated} proposed an AP approach where an ensemble of supervised ML models categorises and prioritises alerts, while analysts leverage their domain knowledge to provide feedback on AP outputs. This feedback is then used to retrain the ML models, reducing repetitive tasks for analysts. Most existing studies follow a similar paradigm to address HAT for AP, relying on periodic model retraining with analyst-labelled alerts~\citep{hore2022towards,kim2022active,liu2022context2vector,hossain2020combating}. 

Although periodic model retraining allows for the integration of human input into AI systems, it offers limited responsiveness to evolving threats due to computational constraints, processing delays, and infrequent update cycles. Moreover, periodic retraining does little to address the brittleness of AI models, which perform badly when encountering out-of-distribution samples \citep{tariq2025a2c}. Even with periodic updates, retrained models may fail to generalise beyond past or current data distributions, leaving them ill-equipped to detect emerging threat patterns or adapt to unforeseen behaviours. Compounding this issue, such models lack the self-awareness to recognise when their predictions are uncertain or likely to fail~\citep{baruwaltowards}. As a result, periodic retraining can create a false sense of robustness while leaving critical gaps in real-world adaptability and reliability.

\subsubsection{Learning to defer (L2D)} \label{L2D}
L2D \citep{madras2018predict} and learning to reject (L2R) \citep{hendrickx2024machine} are frameworks designed to enable ML models to manage uncertain or complex cases. L2R introduces a \textit{rejector}, allowing ML models to refrain from making predictions when encountering uncertain inputs or those beyond their training data. L2D builds upon L2R by not only identifying such cases but actively deferring ambiguous decisions to human experts, who leverage domain knowledge and contextual understanding to make informed decisions~\citep{tariq2025a2c, baruwaltowards}. Some works have extended L2D to include multiple experts~\citep{keswani2021towards,tailor2024learning}, enhancing its versatility. Learning to Complement (L2C) is a variant of L2D that aims to allocate challenging instances to AI when they are difficult for humans, and assign complex instances to humans where AI struggles \citep{wilder2020learning}. To the best of our knowledge, L2D and its variants have yet to be applied to support HAT for AP.

A key limitation of existing L2D setups is their static nature~\citep{joshi2022learning}. Typically, L2D employs two supervised models: a \textit{predictive model} that makes decisions based on input data, and a \textit{deferral model} that determines whether to trust the predictor or defer to a human expert~\citep{liu2019deep,verma2022calibrated}. However, once trained, the deferral model does not adapt to the predictive model’s errors or to human feedback, neglecting the dynamic and evolving nature of real-world decision-making. This lack of adaptability restricts its ability to adjust its deferral policy in real time. In the context of AP in SOCs, such rigidity can be  detrimental, as it hampers the system's capacity to respond properly to evolving threats and undermines effective AP.

Some L2D variants incorporate a RL agent as the deferral model to enhance adaptability ~\citep{balazadeh2022learning, straitouri2021reinforcement,joshi2022learning}. For instance,  \cite{straitouri2021reinforcement} used an RL-based deferral model to determine
which tasks are handled by the AI and which are passed to the human in a car driving task. The RL agent is trained on historical data of both humans and the predictor actions to learn a deferral policy. But, without real-time interaction between human and the RL agent, the L2D model cannot dynamically adapt to evolving contexts, limiting its effectiveness in fast-changing environments.

\subsubsection{Reinforcement learning from human feedback (RLHF)} \label{RLHF}
RLHF is an emerging approach that guides the learning process of RL agents using human feedback, rather than relying on a predefined reward function. This approach holds significant potential for enhancing the adaptability and performance of AI systems, ensuring better alignment with human preferences~\citep{kaufmann2024survey}. Initially developed for training robots to interact with real-world environments, RLHF has since been applied to complex tasks such as Atari games~\citep{christiano2017deep}, recommendation systems~\citep{shuvo2022home,solinas2021peak}, large language models (LLMs)~\citep{zhu2023principled,ouyang2022training}, and image generation~\citep{lee2023aligning}, where human guidance enables RL agents to adapt their decisions in diverse scenarios. An extensive review and discussion on RLHF is provided in~\citep{kaufmann2024survey}.

To the best of our knowledge, only one study has applied RLHF in the AP context.~\cite{wang2024combating} proposed an active learning framework that leverages RLHF, where the RL agent selects the top-ranked, potentially high-risk alerts for analyst validation, with initial priorities assigned by an unsupervised model. This approach incorporates RL in place of traditional supervised learning within the active learning framework, allowing it to dynamically integrate analyst feedback and continuously optimise the active learning's query policy.

While~\cite{wang2024combating} integrate RLHF into active learning for alert selection, their framework lacks the core principle of L2D, i.e., the ability to defer uncertain and unknown decisions to a human expert, which enhances adaptability to novel and evolving threats. Our work fills this gap by introducing L2DHF, an adaptive L2D framework built on RLHF, that facilitates real-time HAT for AP.

\section{Learning to defer with human feedback (L2DHF)} \label{sec:method}

This section introduces the L2DHF framework, as depicted in Figure \ref{PAI-DRLHF}. L2DHF operates through a multi-step process involving predictive AI, AVAR, and a DRL agent operating within the DRLHF framework, facilitating human-AI collaboration. We assume that alerts generated within the SOC are considered for AP at distinct time steps. At each time step, the predictive AI first prioritises incoming alerts. These prioritised alerts are then cross-checked against analyst-validated alerts stored in AVAR from previous steps, ensuring only new or unseen alerts are forwarded to the DRL agent for further evaluation. Previously seen alerts are filtered out. If the predictive AI assigns a priority that matches the AVAR's priority, it is accepted; otherwise, the predictive AI’s priority is overridden by the validated priority provided by AVAR. This filtering process streamlines the AP workflow. The remaining alerts are forwarded to the DRL agent, which functions as an adaptive and  dynamic deferral model. The DRL agent determines whether to defer alerts with potentially inaccurate priorities to human analysts or accept the predictive AI's priorities. Analysts, leveraging their domain knowledge and expertise, validate or adjust the priorities of the deferred alerts, providing feedback to the DRL agent. This continuous feedback helps the DRL agent refine its deferral policy over time, enhancing its ability to defer appropriately to human analysts.

 \subsection{Predictive AI model} \label{sec:PAI}
 
 The predictive AI model is tasked with prioritising alerts. However, its prioritisation may be inaccurate due to several factors, including biased training data, the presence of uncertain or novel alerts that deviate from historical patterns, the model's limited ability to adapt to evolving threats, and challenges in incorporating contextual information. The predictive AI model can be implemented using any supervised learning technique, such as neural networks, random forests, or decision trees, and can also benefit from ensemble methods to improve robustness and accuracy. 

\subsection{Analyst-validated alert repository (AVAR)} \label{sec:AVAR}
AVAR stores alerts that have been validated by analysts over time. At each time step, the DRL agent defers certain alerts to analysts for validation. Analysts either revise the priority assigned by the predictive AI or accept it as is. Since these priorities are validated by human experts, they serve as valuable references for future decisions. This process also allows the system to identify and remove duplicate alerts in subsequent steps, reducing redundancy and easing the workload for both the DRL agent and analysts. By excluding previously validated alerts, AVAR ensures that only new and unseen alerts are forwarded to the DRL agent for further evaluation, thus maintaining the efficiency of the L2DHF framework.

Before alerts prioritised by the predictive AI are forwarded to the DRL agent, they are compared against those stored in AVAR based on their features.  If a match is found, the alert's priority is either retained or adjusted according to the corresponding entry in AVAR. Matched alerts are then excluded from those passed to the DRL agent. 

Furthermore, the alerts stored in AVAR play a key role in shaping elements of the DRL agent's state, supporting its deferral decisions. To facilitate this, AVAR organises analyst-validated alerts into separate storages based on their assigned priority by the analyst. For instance, if alerts are prioritised into categories such as critical, high, medium, low, and normal, AVAR maintains five corresponding category storages and stores analyst-validated alerts in the storage that matches each category.  The use of these category-specific storages in constructing the DRL's state is detailed in Section \ref{sec:state}.

To keep the predictive AI up to date while managing the growing size of AVAR, a cyclical retraining strategy can be employed using the accumulated analyst-validated alerts. Over time, these alerts serve as high-quality training data for regularly  retraining the predictive AI, enhancing its ability to prioritise based on the most recent and relevant feedback. After each retraining cycle, the AVAR can be pruned by removing older, previously used alerts, thereby maintaining scalability and responsiveness. This approach ensures the predictive AI remains adaptive to analyst feedback and evolving threat patterns while keeping the AVAR streamlined.

\subsection{Deep reinforcement learning from human feedback (DRLHF)}  \label{sec:DRLHF}

Through iterative interactions, the DRL agent reviews the alerts prioritised by the predictive AI and either (i) accepts the assigned priorities, or (ii) defers alerts that may be misprioritised to human analysts. Analysts then validate or adjust these priorities based on their domain knowledge. This feedback is used to update the DRL agent's deferral policy. Over time, the continuous feedback loop enables the DRL agent to adapt and optimise its deferral policy, leading to progressively improved performance.

In brief, a DRL agent learns a policy that maximises cumulative rewards through interactions with its environment over time. At each step, the agent:
\begin{itemize}[nosep]
    \item Observes a state $s \in S$, where $S$ is the set of possible states, representing the environment's observable condition.
    \item Chooses an action $a \in A$, from the action set $A$, representing the available choices. 
    \item Receives a reward $r(s, a)$ after taking action $a$ in state $s$.
    \item Transitions to the next state $s' \in S$. 
\end{itemize}

The objective is to learn a policy $\pi$ that maximises the expected cumulative discounted reward:
\[
\mathbb{E}_{\pi} \left[ \sum_{k=0}^{\infty} \gamma^k r^k(s,a) \right]
\]

\noindent where $\gamma \in [0, 1)$ is the discount factor, which determines the relative importance of future rewards and $k \in \mathbb{N}_0$ is an index denoting the number of steps from the current state. The policy $\pi$ or the associated value function is 
approximated using deep neural networks \citep{ravichandiran2020deep}.

The key components of the DRL framework: state, action, and reward, are described below.

\subsubsection{State} \label{sec:state}

The state is represented as a vector of elements for each alert, providing crucial information to guide the DRL agent's deferral decisions. The first element in this vector is the priority assigned by the predictive AI. For
example, this priority can be classified into categories such as critical, high, medium, low and normal. These can be mapped to numerical values (e.g., 4 for critical to 0 for normal) to support efficient analysis, facilitate comparison, and enable seamless integration with ML models. These mappings are flexible and can be adjusted depending on the problem context.

Subsequent state elements are derived from alert features relevant to determining alert priority. Feature extraction techniques can be applied to enhance the quality and relevance of these features. Each alert feature is then compared against the features of analyst-validated alerts stored in the AVAR category storages. If a match is found in a specific category storage, the corresponding state element is assigned that category’s priority. If no match is found, or if the feature matches alerts across multiple categories, the state element is assigned a value of 10 to indicate ambiguity. 

Another state element is computed by calculating the average Euclidean distance between the current alert and the stored alerts in each of the AVAR category storages. The state element is then assigned the priority of the category with the smallest unique average distance. If two or more categories exhibit the same minimum distance, the state element is assigned a value of 10 (unknown). 

The final state element captures the alert's transition status. Initially, it is set to 10 (unknown) for all alerts. When an alert is deferred to the analyst, this element is updated to reflect the priority assigned by the analyst. This update models the alert's current and next state, allowing the DRL agent to improve its learning and refine its decisions.

\begin{figure*}[t!]
  \centering
  \includegraphics[width=0.78\linewidth]{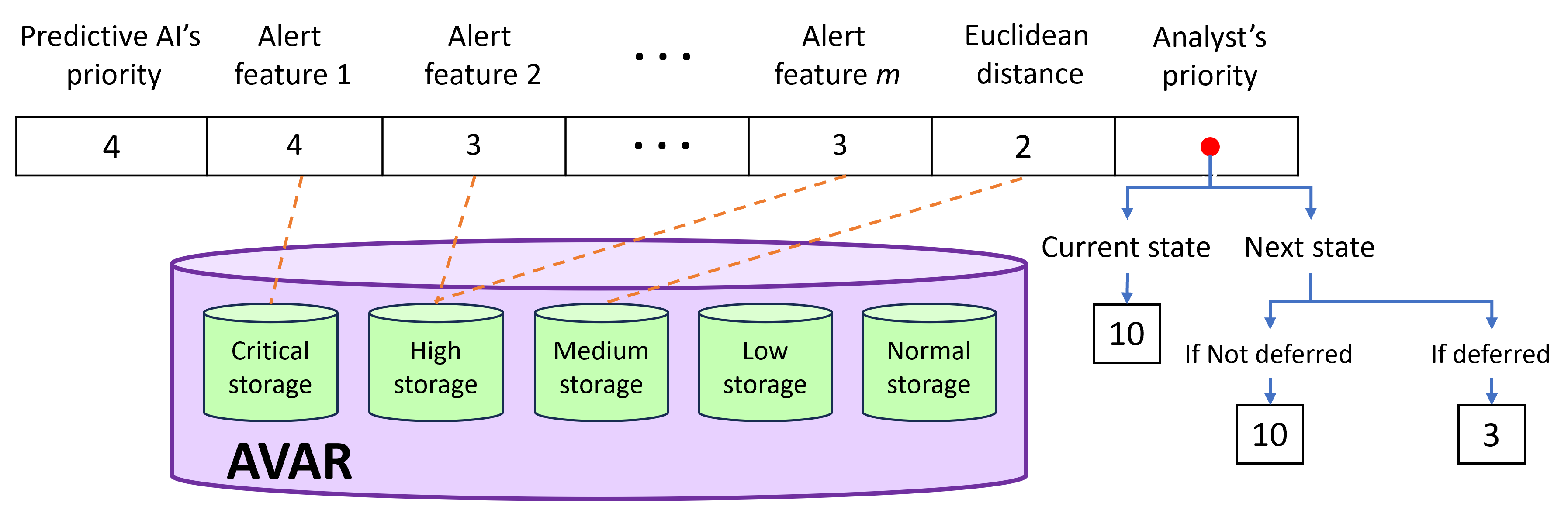}
  \caption{An illustrative example of state elements of DRL.}
  \label{state}
\end{figure*}

Figure \ref{state} provides an illustrative example of the DRL state representation for each alert. In this example, the predictive AI assigns a priority level of "critical" (mapped to the value 4). The alert consists of $m$ features. The value of feature 1 matches that of feature 1 in alerts stored in the critical storage of AVAR, so its state element is assigned 4. The values of features 2 and $m$ match the corresponding features in alerts stored in the high storage, and their corresponding state elements are assigned 3. Also, the alert has the minimum average Euclidean distance to the alerts in medium storage, resulting in a state element value of 2. The final element captures the alert's transition status. If the alert has not yet been processed by the DRL agent, it is in its current state, so this element is set to 10. Once the DRL agent processes the alert, it transitions to its next state and the value of this element is updated. If the alert is deferred, the element takes the analyst-assigned priority (e.g., 3 for high); otherwise, it remains 10.

This state structure is carefully designed to balance the inclusion of informative and learnable elements, contributing to the DRL agent’s decision-making process. The predictive AI’s priority serves as an initial indicator of alert priority. Elements based on similarity between alert features and analyst-validated alerts in AVAR storages incorporate historical expert decisions, enabling the agent to detect potential misprioritisations. The Euclidean distance element offers a complementary, integrated measure of similarity to alerts in AVAR storages. Finally, the inclusion of the transition status further allows the DRL agent to adjust its policy based on how past deferral decisions evolved. This modular design is adaptable to different operational contexts and can be extended or adjusted as needed. 

\subsubsection{Action} \label{sec:action}

The action is represented as a binary variable, where 1 denotes deferring the alert to an analyst, and 0 indicates accepting the predictive AI's assigned priority. An action value of 1 suggests that the alert may have been misprioritised by the predictive AI and requires expert validation, while a value of 0 reflects confidence in the predictive AI's assessment, allowing the system to proceed without human intervention.

\subsubsection{Reward} \label{sec:reward}

The reward quantifies the effectiveness of the DRL agent's deferral decisions based on the analyst's feedback regarding the alert's priority. It is structured around five priority categories: critical, high, medium, low, and normal, which can be tailored to suite the specific alert type and problem context. Eq.(\ref{new_reward}) presents the reward formula:
\begin{equation}
\footnotesize
Reward=
\begin{cases}
z+w, \hspace{0.3cm} p^{\text{AI}} \neq 
 p^{\text{analyst}}, p^{\text{analyst}} \hspace{0.07cm} \text{is Critical} \\
z+h, \hspace{0.3cm} p^{\text{AI}} \neq 
 p^{\text{analyst}}, p^{\text{analyst}} \hspace{0.07cm} \text{is High} \\
z+g, \hspace{0.3cm} p^{\text{AI}} \neq 
 p^{\text{analyst}}, p^{\text{analyst}} \hspace{0.07cm} \text{is Medium} \\
z+f, \hspace{0.25cm} p^{\text{AI}} \neq 
 p^{\text{analyst}}, p^{\text{analyst}} \hspace{0.07cm} \text{is Low} \\
z, \hspace{0.75cm} p^{\text{AI}} \neq 
 p^{\text{analyst}}, p^{\text{analyst}} \hspace{0.07cm} \text{is Normal} \\
-q, \hspace{0.48cm} p^{\text{AI}} = 
 p^{\text{analyst}} \\
0, \hspace{0.72cm} \text{not deferred to the analyst} \\

\end{cases}
\hfill
\begin{aligned}
f<g<h<w\\
z,q,f,g,h,w>0
\end{aligned}
\label{new_reward}
\end{equation}

In Eq.(\ref{new_reward}), $p^{\text{AI}}$ and $p^{\text{analyst}}$ denote the alert priority assigned by the predictive AI and the analyst, respectively. The DRL agent receives a positive reward ($z, z>0$) for deferring inaccurate-priority alerts ($p^{\text{AI}} \neq 
 p^{\text{analyst}}$) and a negative reward ($-q, q>0$) for deferring accurate-priority alerts ($p^{\text{AI}} = 
 p^{\text{analyst}}$) to the analyst. We also introduce additional parameters ($f, g, h, w$,
where $f<g<h<w$) to reward the DRL agent more if it defers an inaccurately prioritised alert with higher criticality to the analyst for correction. 
DRL gets a reward of 0 if it decides not to defer an alert.

The DRL agent checks each alert in the received batch individually,
applying the corresponding state, action, and reward for it. Algorithm \ref{alg:L2DHF} presents the proposed L2DHF framework.

\begin{algorithm}[ht!]
\caption{L2DHF: Learning to Defer with Human Feedback framework.}
\label{alg:L2DHF}
\footnotesize
\KwIn{Raw alerts $\mathcal{L} = \{l_1, l_2, \dots, l_n\}$ at time step $t$}
\KwOut{Prioritised alerts $\mathcal{P}$}
\BlankLine

\ForEach{time step $t$}{
    \tcc{Stage 1: Predictive AI Initial Prioritisation}
    Compute initial priorities $p_i^{\text{AI}}$ for each alert $l_i \in \mathcal{L}$ using the predictive AI\\

    \tcc{Stage 2: AVAR Filtering}
    \ForEach{$l_i \in \mathcal{L}$}{
        \If{$l_i \in \text{AVAR}$}{
            Retrieve AVAR's validated priority $p_i^{\text{AVAR}}$\\
            \eIf{$p_i^{\text{AI}} = p_i^{\text{AVAR}}$}{
                Accept $p_i^{\text{AI}}$ as final priority for $l_i$\\
            }{
                Override $p_i^{\text{AI}}$ with $p_i^{\text{AVAR}}$
            }
        }
        \Else{
            Send $l_i$ to DRLHF module for refinement
        }
    }

    \tcc{Stage 3: DRLHF Priority Improvement}
    \ForEach{$l_i$ sent to DRLHF}{
        Construct DRL state vector $s_i$ for $l_i$\\
        Choose DRL action $a_i$ for $l_i$ where $a_i \in \{\texttt{accept} \ p_i^{\text{AI}}, \texttt{defer} \ l_i\}$\\
        \If{$a_i = \texttt{accept} \ p_i^{\text{AI}}$}{
            Set $p_i^{\text{final}} \gets p_i^{\text{AI}}$\\
           
        }
        \ElseIf{$a_i = \texttt{defer} \ l_i$}{
            Request analyst-validated priority $p_i^{\text{analyst}}$ for alert $l_i$\\
            Set $p_i^{\text{final}} \gets p_i^{\text{analyst}}$\\
            
        } Compute reward $r_i$ using Eq.(\ref{new_reward})\\
        Transition to next state $s'_i$
    } 
}
\Return{$\mathcal{P} = \{(l_i, p_i^{\text{final}}) \mid l_i \in \mathcal{L}\}$}
\end{algorithm}

\section{L2DHF implementation} \label{sec:implement}

This section presents the implementation details of the L2DHF framework, providing a comprehensive overview of its three core components: the predictive AI model responsible for initial alert prioritisation, AVAR that stores analyst-validated alerts, and DRLHF that refines alert priorities through analyst-guided learning. Details on how the analyst is modelled in our experiments are also provided.

\subsection{Predictive AI model} \label{sec:PAI_implement}

\begin{figure}[ht!]
	\centering
	\begin{subfigure}[b]{0.48\linewidth}
		\includegraphics[width=1\linewidth]{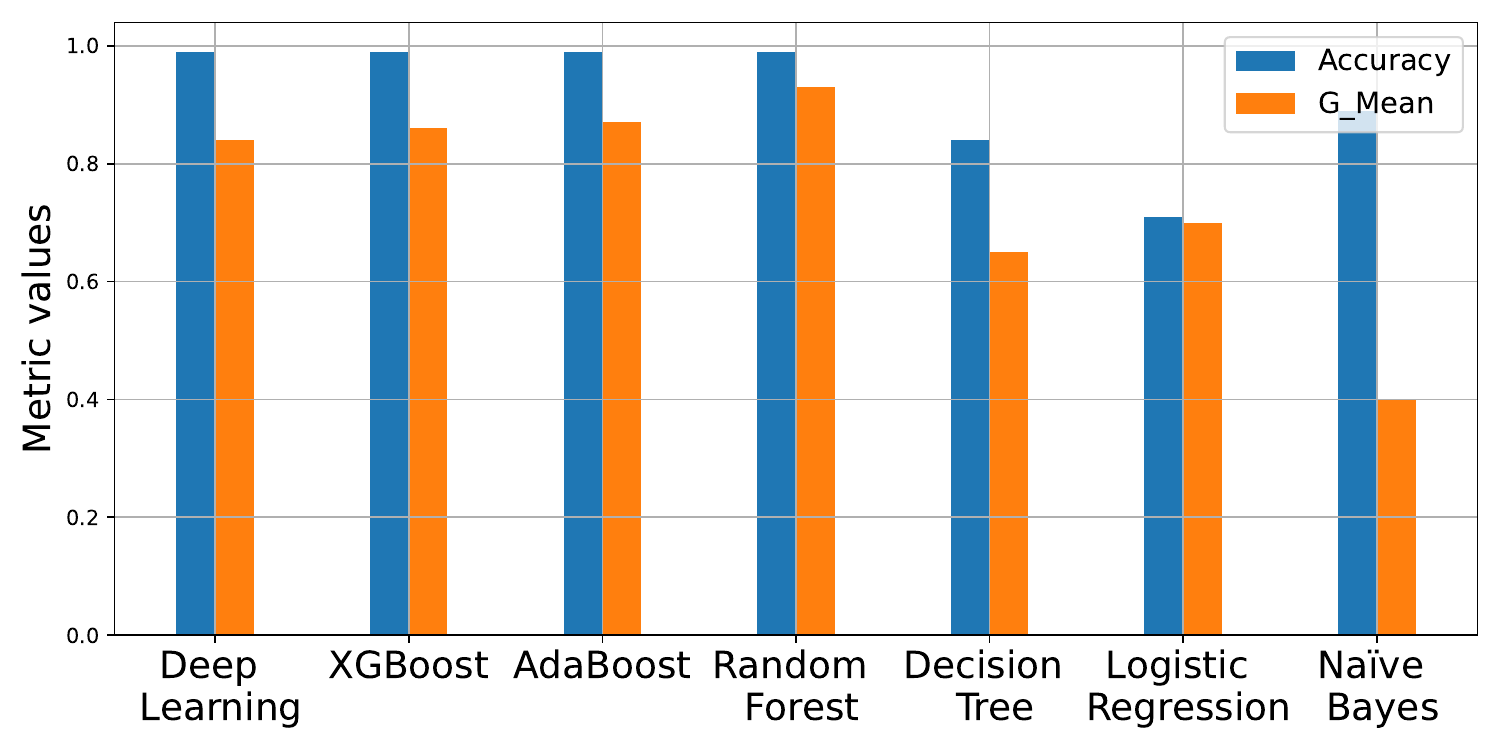}
		\caption{\footnotesize   UNSW-NB15. }\label{fig:classifierUNSW}
	\end{subfigure}
	\begin{subfigure}[b]{.48\linewidth}
		\includegraphics[width=1\linewidth]{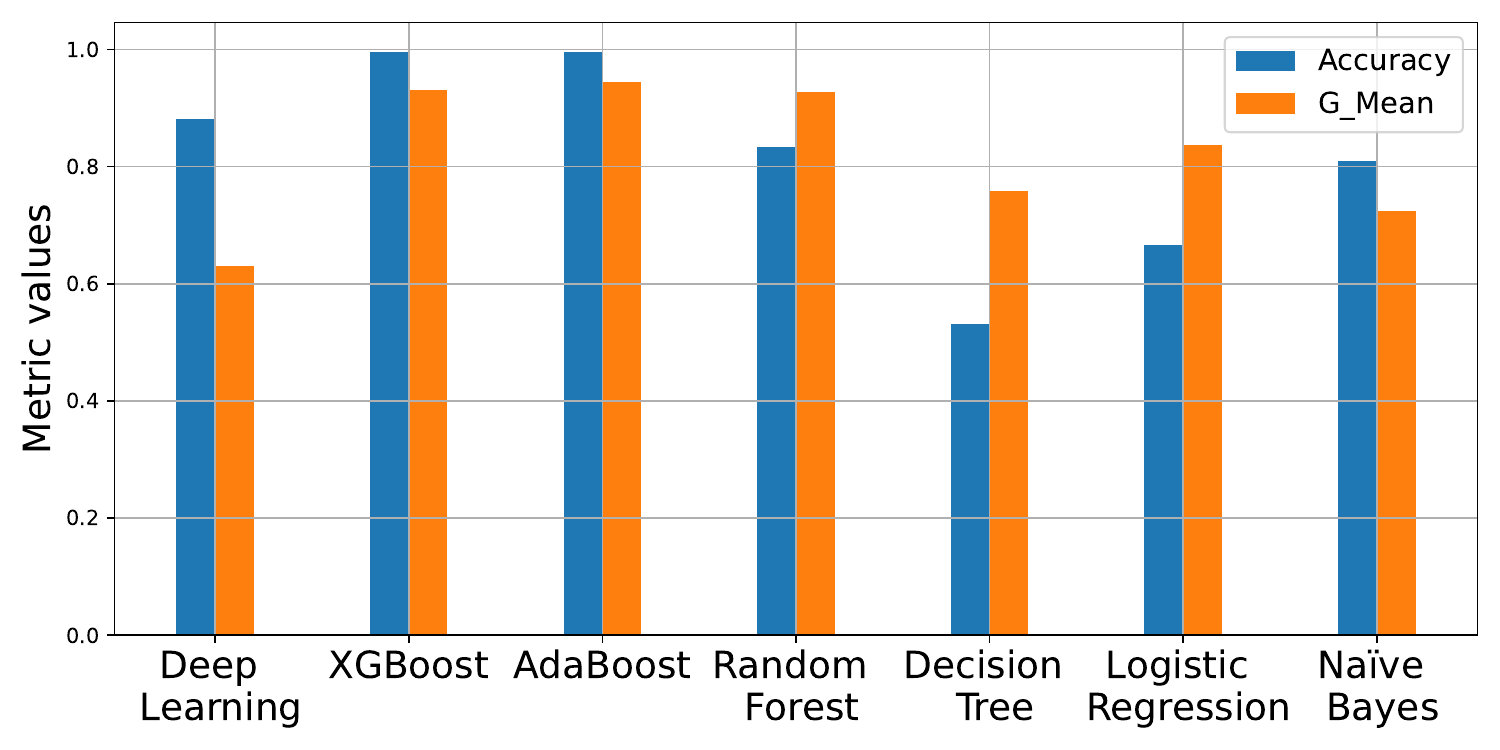}
		\caption{ \footnotesize CICIDS2017.}\label{fig:classifierCIC}
	\end{subfigure}
	\caption{Evaluation of classifier performance across accuracy and G-mean metrics.}
	\label{fig:classifiers}
\end{figure}

The predictive AI was built using an ensemble of classifiers. To achieve this, we initially fine-tuned seven supervised models (Random Forest, Deep Learning, Decision Tree, XGBoost, N\text{\"a}ive Bayes, AdaBoost, and Logistic Regression) using the Optuna hyperparameter optimisation framework~\citep{shekhar2021comparative}. Each model was then trained and evaluated using stratified 10-fold cross-validation over two metrics: accuracy and G-mean. While accuracy reflects the models’ overall effectiveness in AP, G-mean captures their ability to handle class imbalance, an important consideration in real-world security datasets. Based on performance across these metrics, we selected the top four models for each dataset. As shown in Figure \ref{fig:classifiers}, for UNSW-NB15, Random Forest, Deep Learning, XGBoost, and AdaBoost demonstrated consistently strong performance across both metrics and were chosen for the ensemble. For CICIDS2017, Random Forest, XGBoost, AdaBoost, and N\text{\"a}ive Bayes were chosen. Although Deep Learning exhibited slightly higher accuracy than Random Forest and N\text{\"a}ive Bayes, its low G-mean indicated weaker performance on imbalanced datasets, leading to its exclusion for this dataset.

The ensemble model was used to categorise alerts into five severity levels based on the Common Vulnerability Scoring System (CVSS): \textit{critical, high, medium, low} and \textit{normal}. Severity levels are commonly used for AP in existing studies \citep{bicudo2024statistical,salman2020modeling,tripathi2011prioritization}.

\subsection{AVAR} \label{sec:AVAR_implement}
AVAR was implemented as outlined in Section \ref{sec:AVAR}, with its category storages organised  according to the five CVSS severity levels. As our experiments span a relatively short period (12 weeks, as noted in Section \ref{sec:OperatCond}), we did not retrain the predictive AI model or refresh the AVAR over this period. This allowed for a focused evaluation of the deferral mechanism under stable conditions, without added variability from model updates.

\subsection{DRLHF} \label{sec:DRLHF_implement}

The DRLHF implementation was structured around the core components of DRL: state, action, and reward. The implementation of the action component follows the design described in Section \ref{sec:action} and is not repeated here. The following details the implementation of the state and reward components. 


\begin{itemize}[nosep]
    \item \textbf{State.} The state was represented as a vector comprising multiple elements, constructed as outlined in Section \ref{sec:state}. Additional implementation details are provided here for the elements derived from alert features, which require further elaboration. Our experiments utilised alerts from two datasets, UNSW-NB15 and CICIDS2017. We applied Principal Component Analysis (PCA) for feature extraction and dimensionality reduction, resulting in 12 principal components per dataset. We evaluated four configurations for selecting these features: (1) the top 3, (2) top 6, (3) top 9, and (4) all 12 PCA components. For each dataset, we empirically determined the optimal number of alert-feature based state elements by assessing the AP accuracy of L2DHF across these configurations. The optimal setup used all 12 PCA features for UNSW-NB15 and the top 6 PCA features for CICIDS2017. These features were then used to construct the corresponding state elements for the DRL agent.
    
    \item \textbf{Reward.} The reward was determined according to Eq.(\ref{new_reward}), using the following parameter values: $q=5, z=1, f=2, g=4, h=6, w=8$. These parameters are configurable and can be adapted to suit different problem contexts. Although fine-tuning these values may optimise performance, it is not central to assessing the effectiveness of L2DHF in comparison with the baseline models.
\end{itemize}

\subsubsection{DRL algorithm and architecture} \label{sec:DRLalgorithmArchitecture}
We employed the Dueling Double Deep Q Network (D3QN) \citep{gok2024dynamic}, a RL architecture well-suited to problems with discrete action spaces. D3QN builds upon the widely used Deep Q Network (DQN) algorithm by integrating Double DQN and Dueling DQN. In doing so, D3QN addresses DQN's issue of overestimating action values, while also speeding up the training process \citep{gok2024dynamic}.

D3QN, like DQN and its variants, employs a neural network to approximate the Q-value,   
$Q(s,a)$, representing the expected return of taking action $a$ in state 
$s$. The network architecture used to implement D3QN consists of an input layer followed by three fully connected (FC) layers (64, 64, and 32 neurons), each employing ReLU activation. The output layer uses a linear activation with dimensionality equal to the action space. The model uses the Adam optimiser with a 0.001 learning rate and the mean squared error (MSE) loss. We also set the size of replay buffer (a memory storing past experiences of DQN for training the neural network model) to 1000 and the batch size (the number of experiences sampled from the replay buffer for training the network) to 64.

For the RL part of D3QN, we set the discount factor to $\gamma = 0.70$ and the greedy parameter to $\epsilon_0 = 0.75$. To ensure Q-value convergence \citep{li2019adaptive}, we adopted a step-decay approach, which gradually decreases the greedy parameter $\epsilon$ over time steps. In early time steps, $\epsilon$ was kept high to promote exploration, then gradually reduced to facilitate exploitation. Eq.(\ref{epsilon}) calculates $\epsilon$ for each time step:

\begin{equation}
	\footnotesize
	\epsilon=\frac{\epsilon_{0}}{1+\text{time step}*d\_\epsilon}
	\label{epsilon}
\end{equation}

\noindent where $d\_\epsilon$ is the decay level for $\epsilon$ and is set to 0.005. We also set a minimum value of $\epsilon_{min} = 0.01$ to prevent $\epsilon$ from falling below this threshold.

We implemented both the D3QN algorithm and the associated alert prioritisation environment from scratch, ensuring they were tailored to the specific requirements and constraints of the AP task.

\subsubsection{Analyst} \label{sec:analyst}
In our implementation, ground truth data was used to simulate the analyst's feedback during interactions with the DRL agent. This proxy is widely adopted in prior work on human-AI decision-making~\citep{mozannar2020consistent,wang2024combating, cao2024reinforcement}. For instance, \cite{wang2024combating} use ground truth data to represent security analysts’ feedback in an anomaly detection-based AP approach that incorporates active learning through RLHF. Similarly, our framework uses ground truth data as the known priority labels of alerts from the datasets to emulate how an analyst would assess the alert’s priority, allowing the DRL agent to receive timely and consistent reward signals.

The interaction between the DRL agent and the analyst was subjected to a limited time budget, reflecting the analyst's constrained availability. While the DRL agent can evaluate all incoming alerts, the analyst may not have sufficient time to review every deferred alert in a given iteration. Following the approach in~\citep{shah2018two}, we assumed that approximately 80\% of an analyst’s time is spent on reviewing alerts, with the remaining 20\% allocated to other tasks such as training and report writing. For instance, if each time step corresponds to one hour, the analyst can devote roughly 48 minutes to reviewing alerts deferred by the DRL agent.

According to MITRE, analysts typically spend several minutes reviewing each alert, though the exact duration varies depending on factors such as SOC policies, the number of analysts, and the volume of alerts~\citep{knerler202211}. Following MITRE~\citep{knerler202211}, we assume that each alert requires a few minutes for review, with more severe alerts requiring additional time. Table  \ref{tab:analyst time} outlines the assumed analyst review durations for each alert category. These time allocations are specific to the problem context and can be adjusted to reflect different operational environments or organisational policies. 

\begin{table*}[h!]
\centering
\scriptsize
  \caption{Analyst's review time for different alert categories.}
  \label{tab:analyst time}
  \setlength{\tabcolsep}{4pt}
  \begin{tabular}{l c c c c c}
    \toprule
    
      Alert category & Critical  &   High & Medium &    Low & Normal \\
      
   \midrule

       Review time (min) & 4.5 & 3.5 & 2 &  1.5  & 1 \\

  \bottomrule
\end{tabular}
\end{table*}

Additionally, our implementation of L2DHF involved only one analyst. To increase the processing of more severe alerts, L2DHF sorts alerts by their predictive AI's priority before the DRLHF phase.

\section{Experimental setup} \label{sec: experimental setup}

\subsection{Datasets} \label{sec:dataset}
We utilised two widely used, publicly available network intrusion datasets to evaluate L2DHF: UNSW-NB15~\citep{moustafa2016evaluation,sarhan2021netflow,moustafa2017novel} and CICIDS2017~\citep{sharafaldin2018toward}. Although other datasets such as KDD-Cup99 \citep{kdd_cup_1999_data_130} and NSL-KDD \citep{tavallaee2009detailed} are available, UNSW-NB15 and CICIDS2017 were chosen as representative benchmarks that sufficiently capture a wide range of typical intrusion scenarios. While additional datasets may offer further insight, these two are considered adequate for demonstrating the performance advantages of L2DHF compared to baseline models. Each dataset contains more than 2 million alerts, offering a robust foundation for evaluation. 

The UNSW-NB15 dataset contains 49 features, both numerical and categorical \citep{ngo2023machine}, including flow features (e.g., source/destination IP addresses and port numbers), basic features (e.g., total duration and source bits per second), and time-related features (e.g., record start time and end time) \citep{moustafa2016evaluation}. Similarly, CICIDS2017 has both numerical and categorical features, including flow features, as well as additional features like duration and total forward inter-arrival time \citep{sharafaldin2018toward}. The total number of features of CICIDS2017 is 79. The complete list of features for UNSW-NB15 and CICIDS2017 can be found in \citep{moustafa2016evaluation} and \citep{rosay2022multi}, respectively.

Alerts were prioritised according to the five CVSS severity levels. For UNSW-NB15, the CVSS values were sourced from the ground truth data. For CICIDS2017, the CVSS values were sourced from \citep{duraz2023cyber}, which provides a summary table of numerical CVSS scores. These scores were then mapped to five categorical levels using the standard defined by National Institute of Standards and Technology (NIST) and CVSS v3.x. \citep{NIST2024}.

Both datasets were pre-processed to ensure data quality and consistency. Following the approach in~\citep{ngo2023machine}, we removed flow-related features such as source and destination IP addresses. Features with a high proportion of missing values were also excluded, and categorical features were transformed into numerical representations using one-hot encoding. To further enhance feature quality and reduce dimensionality, we applied PCA, a widely used technique~\citep{ngo2023machine,moustafa2017novel,zoghi2024unsw}, to extract the most relevant features. Based on the cumulative variance ratio, we selected 12 principal components for both datasets, as this configuration captures over 95\% of the total variance,  ensuring adequate information retention. These PCA-transformed alerts were used as input for the predictive AI model's initial AP. 

Each dataset was divided into two parts: (i) building the ensemble of classifiers and (ii) evaluating L2DHF. For ensemble development, 250,000 samples per dataset (around 10\% of the total data) were used, with 25,000 samples (10\%) for parameter tuning, 175,000 samples (70\%) for training, and 50,000 samples (20\%) for testing the classifiers. Based on the information in Section \ref{sec:OperatCond} regarding the number of time steps and average arrival rate of alerts, the average number of alerts used for evaluating L2DHF was 806,400 in both datasets.


\subsection{Baseline models} \label{sec:BaseModels }

We evaluated L2DHF against three baseline models: (i) the predictive AI ensemble, (ii) DRLHF, and (iii) an L2D model. Details of the DRLHF and predictive AI implementations are provided in Section~\ref{sec:implement}. The L2D model utilises a confidence threshold to determine whether an alert should be deferred to the analyst. With four classifiers in the ensemble, a priority is accepted if at least three classifiers agree on the same value, establishing a confidence threshold of 0.75. If the priority confidence falls below this threshold, the L2D model defers the alert to the analyst; otherwise, it retains the priority assigned by the predictive AI. This approach assumes that the analyst's prediction error is constant, as described by \citep{madras2018predict}. By leveraging ground truth data as the analyst feedback, our model aligns with this assumption.

\subsection{Evaluation metrics} \label{sec:evaluationMetrics}

We evaluated L2DHF and baseline models using the following metrics. Table \ref{tab:metrics_formulas} summarises the formulas and descriptions of the metrics. 

\begin{itemize}[nosep]
    \item \textbf{AP Accuracy:} Measures the model's ability to accurately prioritise alerts across different severity categories. It is defined as the ratio of correctly prioritised alerts to the total number of alerts prioritised.
    
    \item  \textbf{Misprioritisations:} This metric counts all alerts that are incorrectly assigned to severity levels different from their true category. It reflects the model’s overall effectiveness in minimising misprioritisations that could compromise security. For example,  misprioritisation of critical alerts includes cases where critical alerts are mistakenly assigned high, medium, low, or normal priorities. In contrast, false positives and false negatives represent specific types of  misprioritisations:
    false negatives occur when genuine threats—alerts with severity levels of critical, high, medium, or low—are incorrectly categorised as non-threats, i.e., in the normal category. Conversely, false positives are instances where non-threatening alerts (normal category) are mistakenly assigned a threat level of low, medium, high, or critical.

    
    \item \textbf{Unprocessed Alerts and Deferred Alerts by the Deferral Model:} These interrelated metrics assess the deferral model's effectiveness in handling alert volume and analyst workload. The number of unprocessed alerts represents those received for prioritisation but not processed by \textit{the deferral model}, indicating missed opportunities to improve AP; a lower count indicates better performance. In contrast, the number of processed alerts refers to those handled by the deferral model. Moreover, the number of deferred alerts reflects how often the deferral model defers alerts to human analysts, directly impacting their workload.

    \item \textbf{Execution Time:} This metric represents the computational time, in seconds, required for the model to process alerts at each time step.
    Shorter execution times indicate greater efficiency, enabling real-time processing in SOC environments.

%
\end{itemize}

\begin{table}[t!]
\centering
\caption{Formulas and descriptions of evaluation metrics.} \label{tab:metrics_formulas}
\footnotesize
\begin{tabular}{l l l}
\toprule
\textbf{Metric} & \textbf{Formula} & \textbf{Description} \\
\midrule

AP Accuracy  & 

$\text{AP Accuracy}_c = \frac{|\{i \in A_c^p : \hat{p}_i = p_i\}|}{|A_c^p|}$ 
& where $A_c^p$ is the set of alerts prioritised into category $c$ ($c$: critical,  
\\

(per category $c$) & & high, medium, low, normal), $\hat{p}_i$ is the 
 assigned priority, and $p_i$ the   \\

  & & true priority. Averaged over all time steps. \\[1em]

 Misprioritisations & 
 $\text{MP}_c = \sum_{i \in A_c^T : p_i = c \land \hat{p}_i \neq c} 1$
 
& where $A_c^T$ is the set of alerts belonging to category $c$. This measures 
\\

(per category $c$) & & the total number of alerts with category \(c\) but incorrectly prioritised \\

 & & into other categories, summed over all time steps.\\[1em]

Unprocessed Alerts 
&  $
U = |A^{\text{rec}}| - |A^{\text{proc}}|
$ & where $A^{\text{rec}}$ is the set of alerts received, and \(A^{\text{proc}}\) is the subset that 
\\

 (by the deferral model) & &  were processed by the deferral model.\\[1em]

Deferred Alerts  & $
D = \sum_{i \in A^{\text{proc}} \cap A^{\text{defer}}} 1
$
 &
where $A^{\text{defer}}$ is the set of alerts deferred to the analyst. This measures 
\\

  & &  total number of alerts that were both processed by the deferral model  \\

 & &   and then deferred to the analyst for review.\\[1em]

Execution Time & 

$T_{\text{exec}}$ (in seconds)
 &
Measures the model’s runtime to process all alerts in a time step. 
\\


\bottomrule
\end{tabular}

\end{table}

\subsection{System specifications and execution environment} \label{sec:OperatCond}

The models were implemented in Python 3.11 and executed on an Intel(R) Xeon(R) Gold 6148 machine with 65GB of RAM and 12 CPU cores running at 2.40GHz.

The models were run continuously to simulate alert processing over a 12-week period, representing 24/7 SOC operations across 2016 hourly time steps. This duration allowed the system to capture diverse alert characteristics and evolving analyst-system interactions, while providing the DRL agent ample opportunity to learn and converge on optimal decisions. At each hourly time step, the number of incoming raw alerts was modelled using a Poisson distribution, as outlined in \citep{shah2018two,hore2023deep,huang2022radams}. The distribution had an average arrival rate of 400 alerts per hour, reflecting typical real-world SOC conditions, where approximately 10,000 alerts are received daily \citep{van2022deepcase,FireE}. These alerts were then forwarded to the models for prioritisation.

The L2DHF framework is designed to be both scalable and adjustable to diverse operational conditions across different SOCs. While our implementation used a one-hour time step and an average of 10,000 alerts per day, the framework can easily be reconfigured for different time intervals and alert volumes. SOCs with lower or higher traffic can adjust the time step duration or alert arrival rates accordingly, ensuring effective deployment in both smaller-scale SOCs with moderate alert loads and in high-throughput SOCs requiring real-time prioritisation over shorter intervals.

\section{Results} \label{sec: results}
This section presents the results and discusses the key insights. Table \ref{tab:summary} summarises the key findings related to L2DHF performance. The results are presented below with reference to the relevant evaluation metrics.

\begin{table}[ht!]
\centering
\footnotesize
  \caption{Summary of L2DHF performance across all metrics.}
  \label{tab:summary}
  \setlength{\tabcolsep}{1.5pt}
  
  \begin{tabular}{l l }
    \toprule
    
     \textbf{No.} & \multicolumn{1}{c}{\textbf{Description}}  \\

     
    \midrule
    
     \multirow{2}{*}{1} &  L2DHF consistently outperforms L2D, the predictive AI, and DRLHF in terms of AP accuracy across various categories\\ 
   &     and overall.  \\

\midrule

   \multirow{2}{*}{2}& L2DHF reduces the number of unprocessed alerts, thereby enhancing AP accuracy, and reduces the number of deferred alerts, 
    \\

    &  thus decreasing the analyst workload.\\

 \midrule

 \multirow{2}{*}{3}&  
 L2DHF reduces misprioritisations. In particular, it decreases the misprioritisation of top-severity alerts to lower severity
  \\

&    categories, thereby lowering the likelihood of top-severity alerts being overlooked and potential security compromises.\\

\midrule
    
\multirow{2}{*}{4}&  Handling unprocessed alerts typically necessitates increasing the number of analysts. L2DHF, with its superior AP \\

&   accuracy and reduced number of deferred alerts, requires fewer analysts compared to baselines.\\

\midrule

    \multirow{2}{*}{5} & DRLHF prioritises significantly fewer alerts than L2DHF, resulting in a large number of unprocessed alerts. Since DRLHF
 \\

& 
       does not incorporate initial prioritisation from the predictive AI, the unprocessed alerts have no assigned priority. \\

     \midrule

    \multirow{2}{*}{6} & Improvements in the predictive AI can enable L2DHF to process a greater proportion of alerts, leading to higher AP
 \\
& 
    accuracy and overall system performance. \\

     \midrule

7 & L2DHF demonstrates manageable and efficient execution times, making it suitable for real-time AP scenarios.\\

  \bottomrule
\end{tabular}
\end{table}

\subsection{AP accuracy} 
\begin{figure}[!t]
	\centering
	\begin{subfigure}[b]{0.45\linewidth}
		\includegraphics[width=0.95\linewidth]{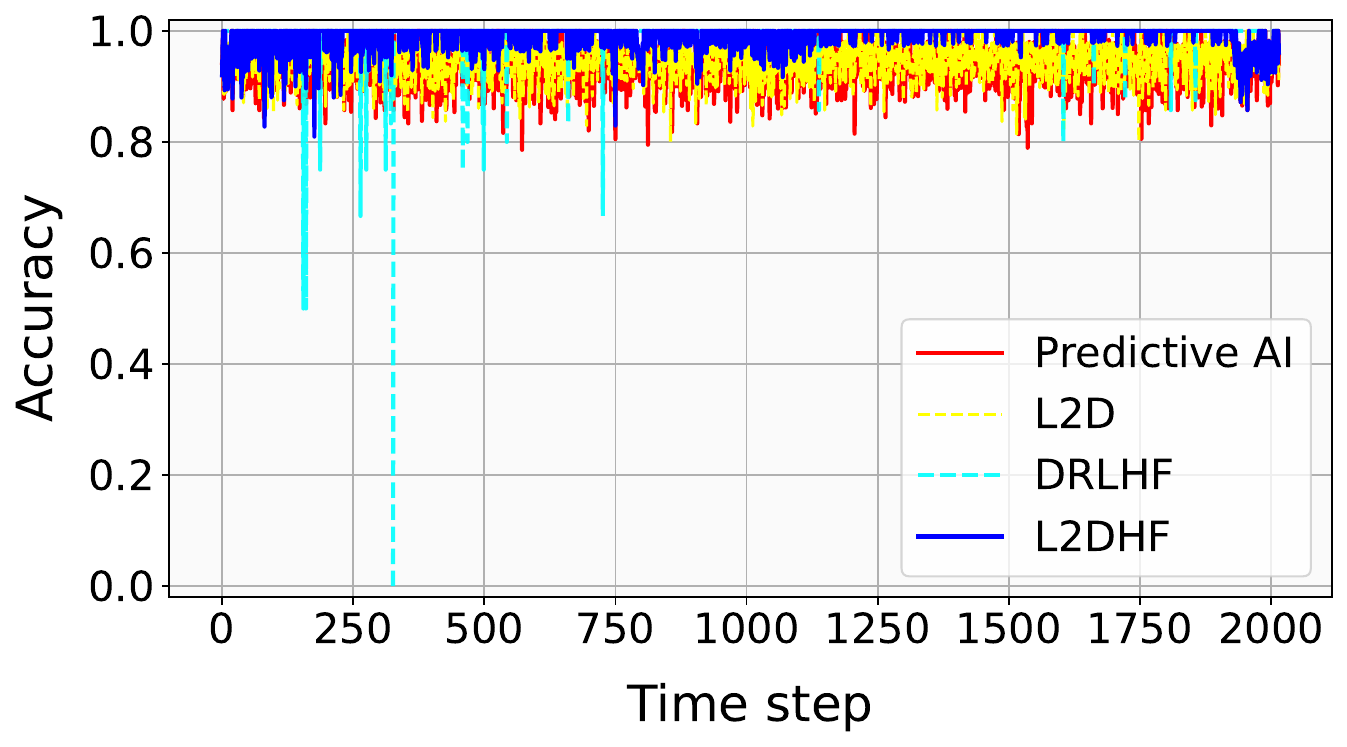}
		\caption{\footnotesize  Accuracy overall (without normal category). }\label{fig:Accuracy_overal_UNSW}
	\end{subfigure}
	\begin{subfigure}[b]{.45\linewidth}
		\includegraphics[width=0.95\linewidth]{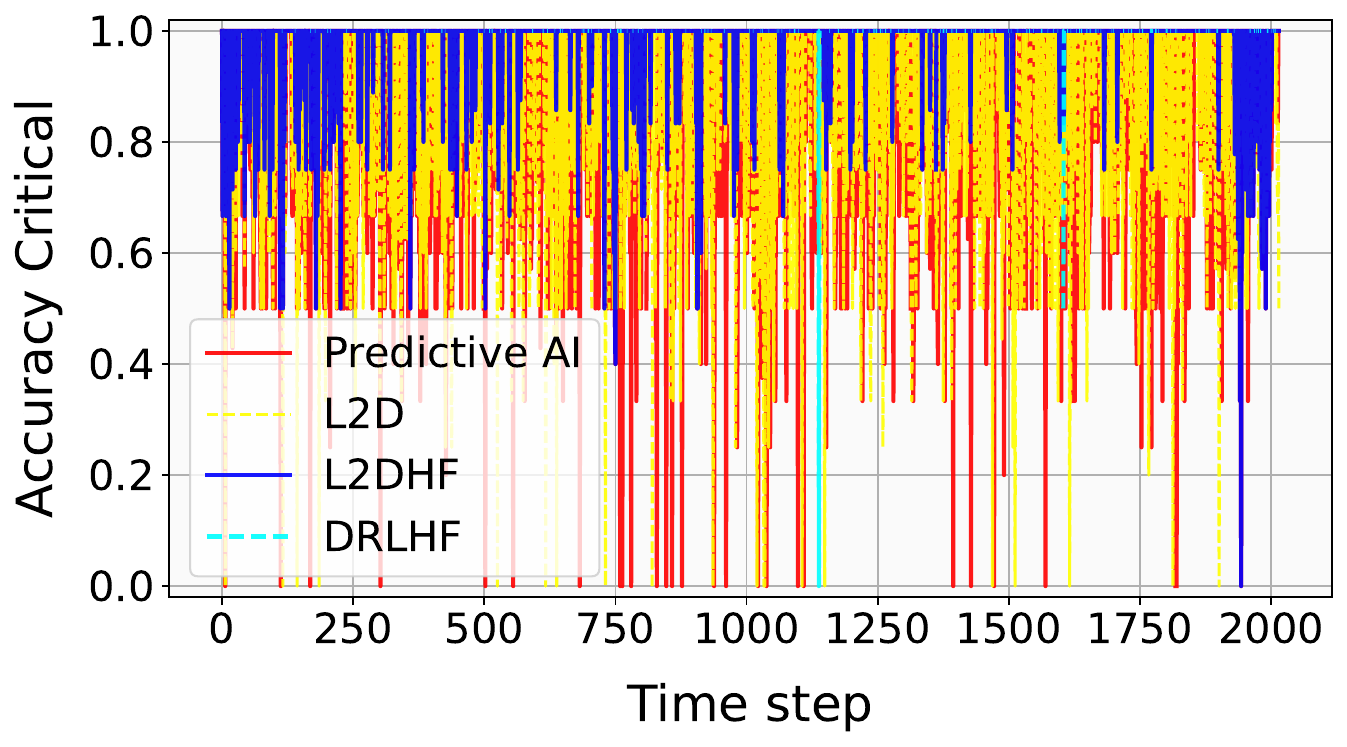}
		\caption{ \footnotesize Accuracy critical category.}\label{fig:Accuracy_Critical_UNSW}
	\end{subfigure}
	\begin{subfigure}[b]{0.45\linewidth}
		\includegraphics[width=0.95\linewidth]{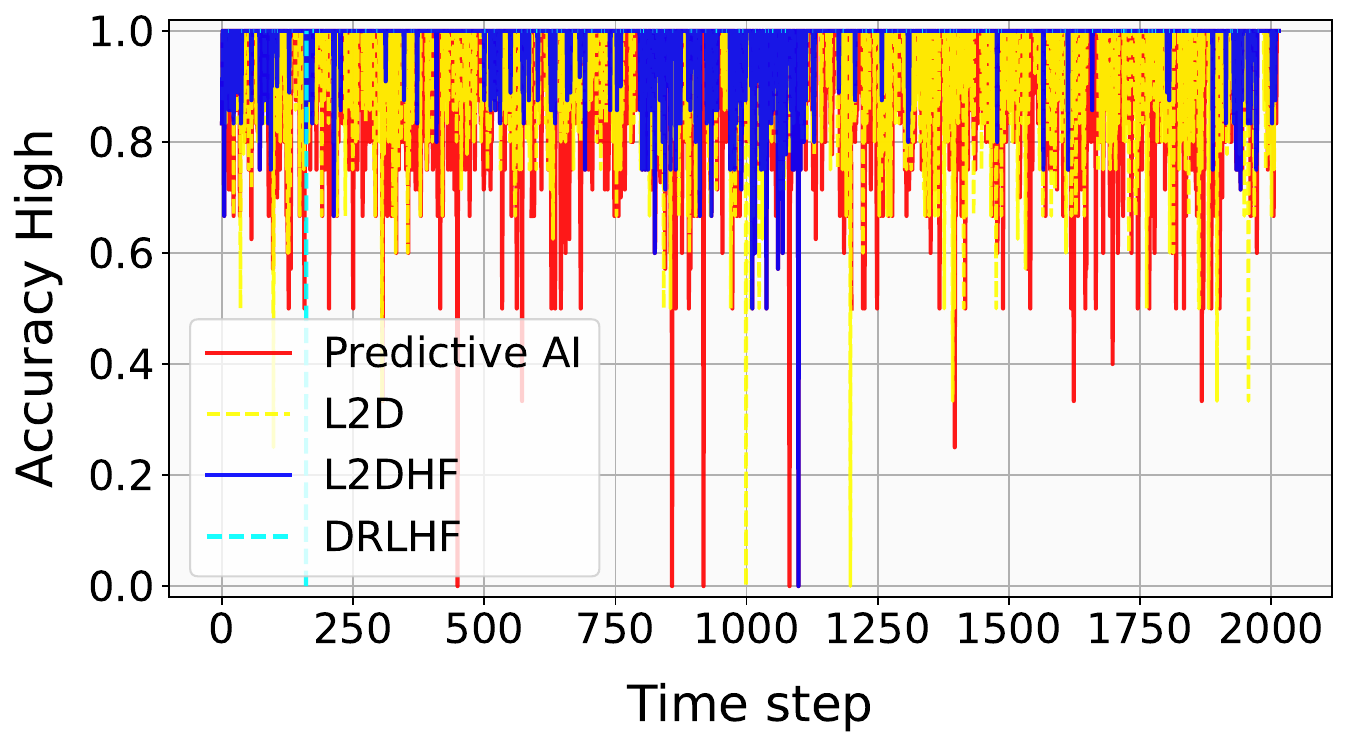}
		\caption{ \footnotesize Accuracy high category.}\label{fig:Accuracy_High_UNSW}
	\end{subfigure}
	\begin{subfigure}[b]{0.45\linewidth}
		\includegraphics[width=0.95\linewidth]{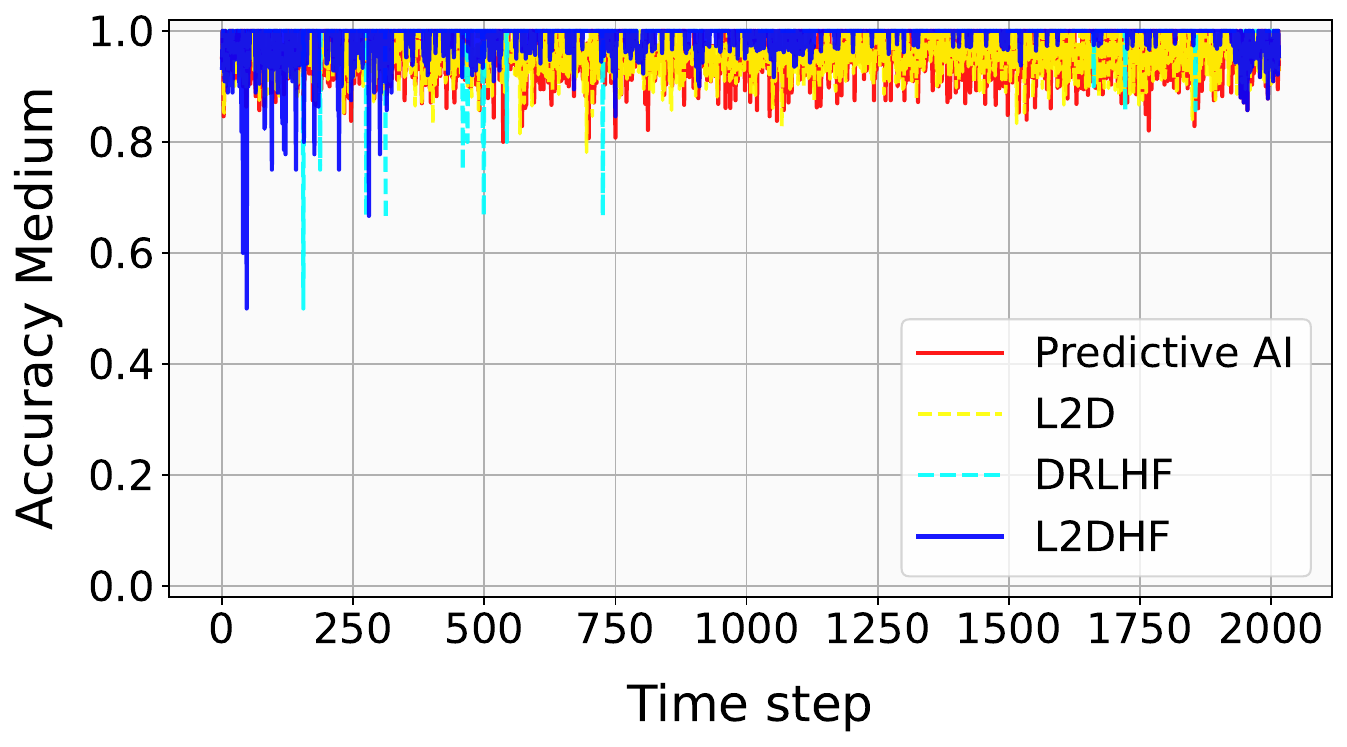}
		\caption{ \footnotesize Accuracy medium category.}\label{fig:Accuracy_Medium_UNSW}
	\end{subfigure}
	\begin{subfigure}[b]{0.45\linewidth}
		\includegraphics[width=0.95\linewidth]{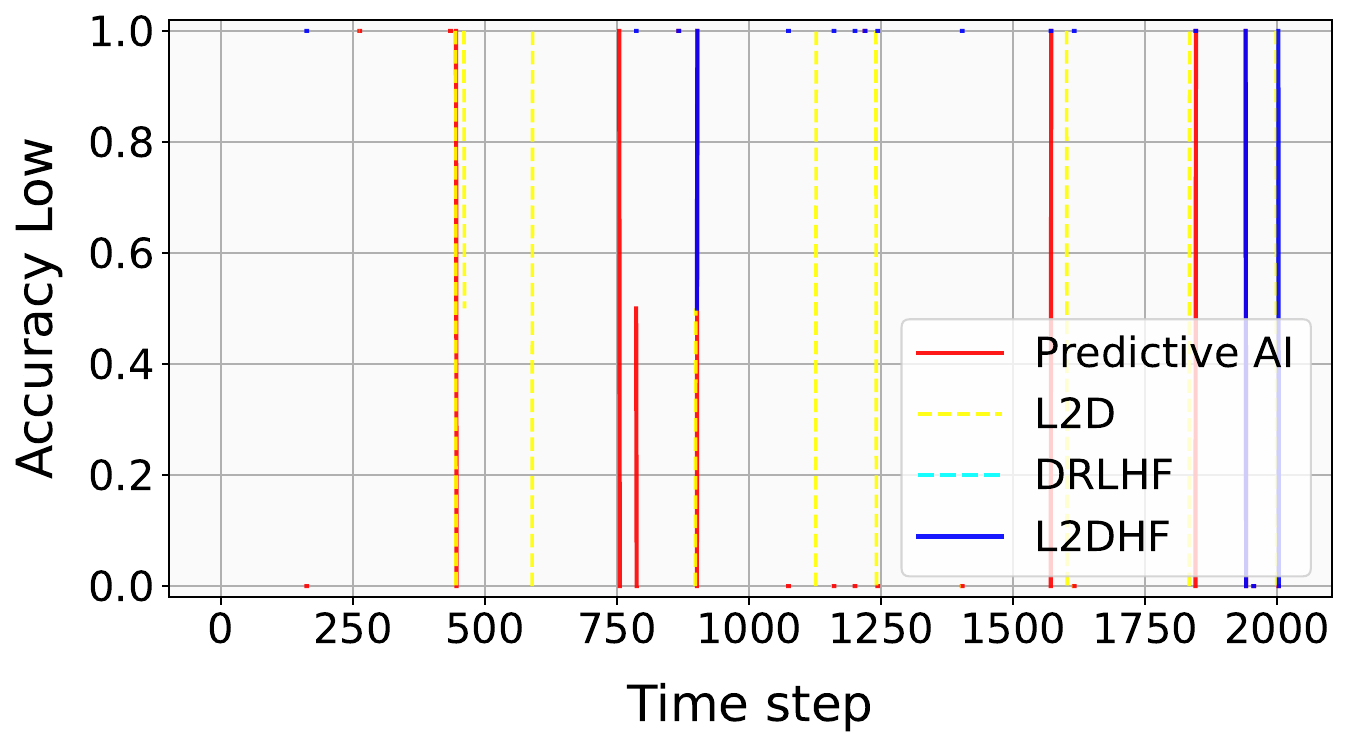}
		\caption{ \footnotesize Accuracy low category.}\label{fig:Accuracy_Low_UNSW}
	\end{subfigure}
	\caption{AP accuracy in categories and overall, UNSW-NB15.}
	\label{fig:accuracy_UNSW-NB15}
\end{figure}

\begin{figure}[ht!]
	\centering
	\begin{subfigure}[b]{0.45\linewidth}
		\includegraphics[width=0.95\linewidth]{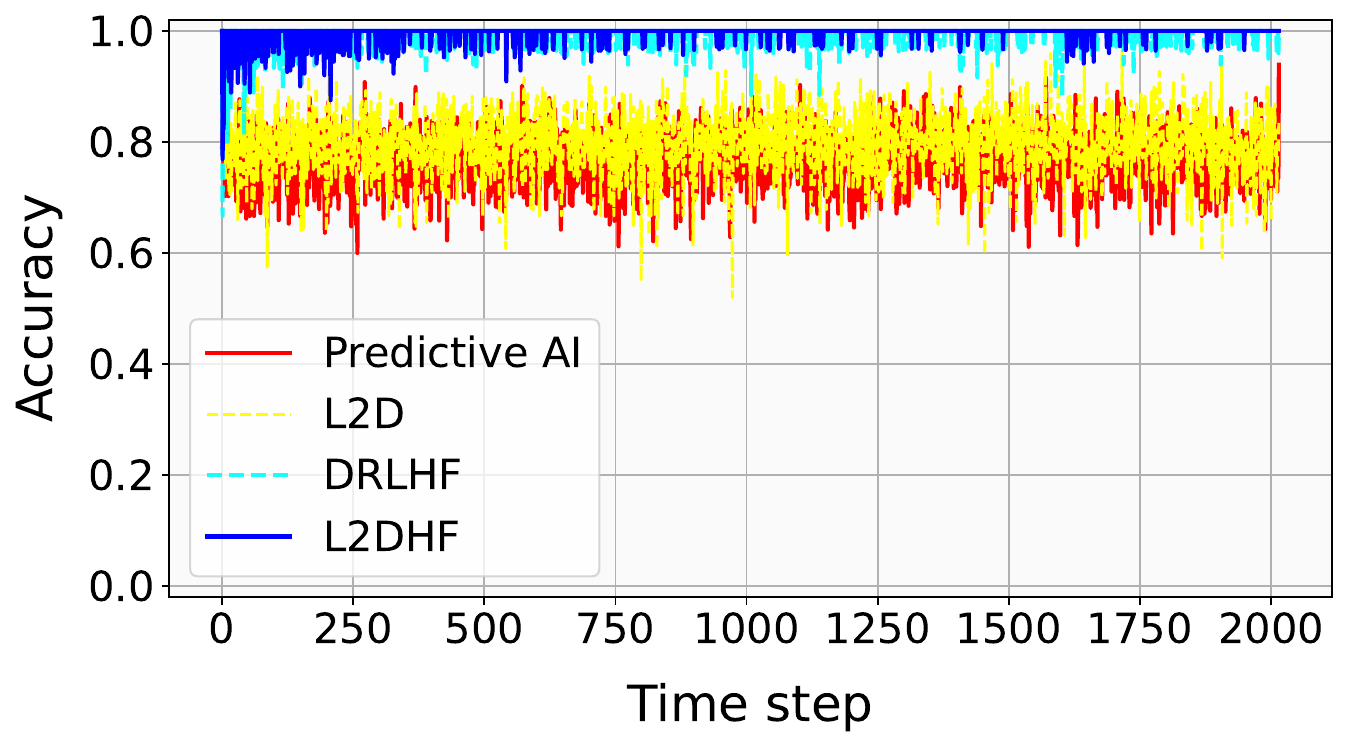}
		\caption{\footnotesize  Accuracy overall (without normal category). }\label{fig:Accuracy_overal_CICIDS}
	\end{subfigure}
	\begin{subfigure}[b]{.45\linewidth}
		\includegraphics[width=0.95\linewidth]{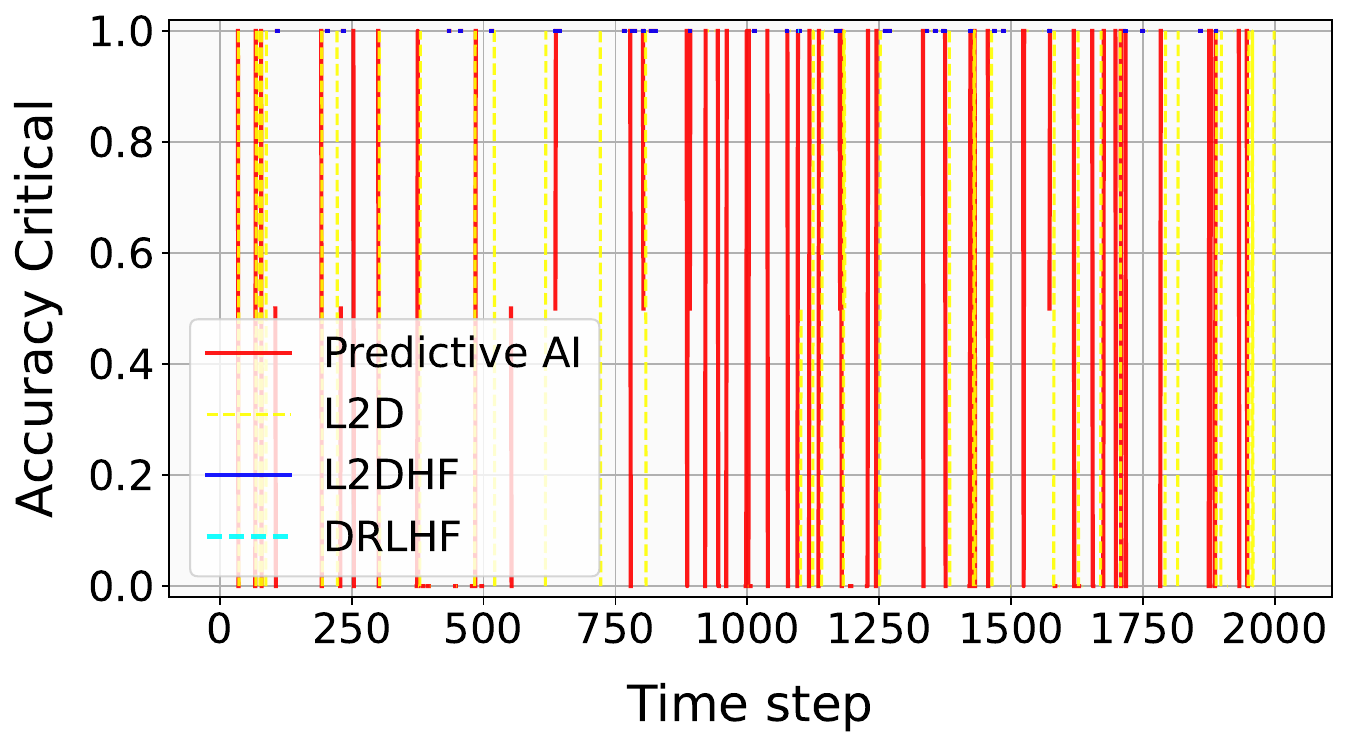}
		\caption{ \footnotesize Accuracy critical category.}\label{fig:Accuracy_Critical_CICIDS}
	\end{subfigure}
	\begin{subfigure}[b]{0.45\linewidth}
		\includegraphics[width=0.95\linewidth]{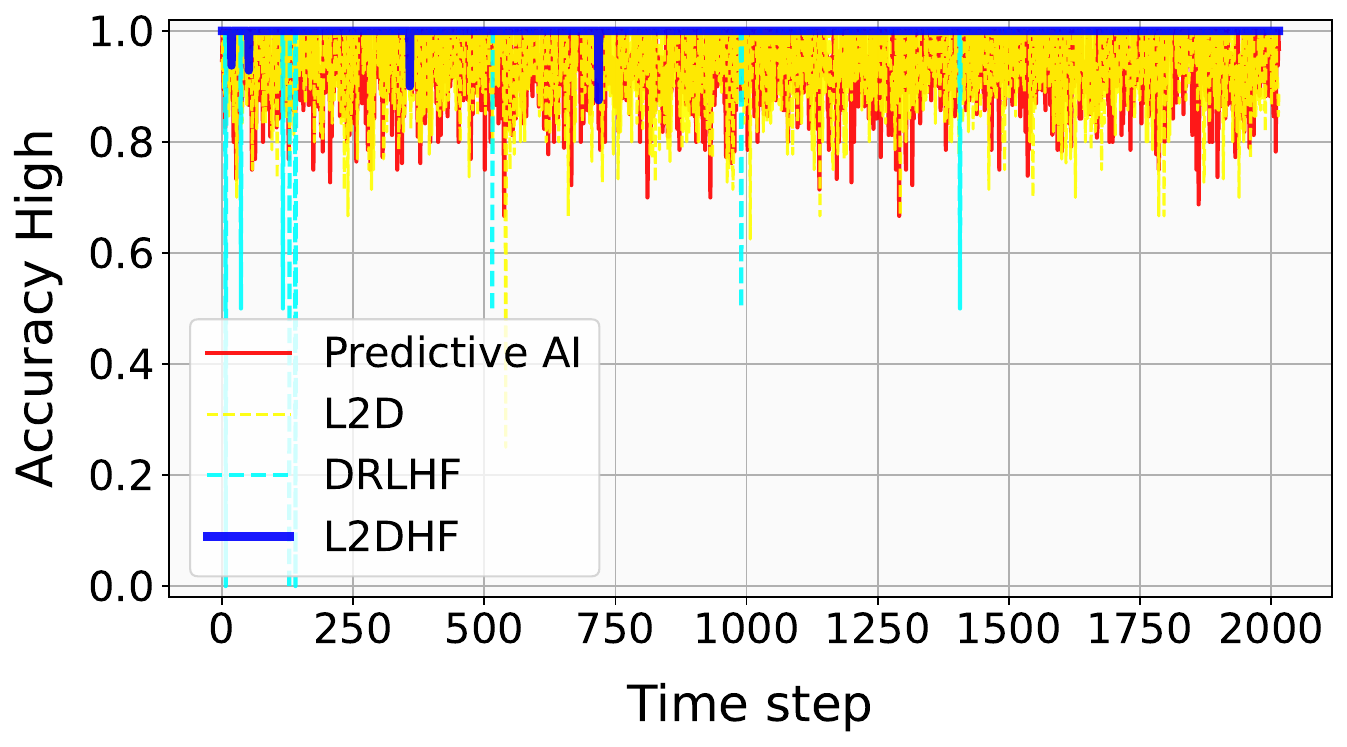}
		\caption{ \footnotesize Accuracy high category.}\label{fig:Accuracy_High_CICIDS}
	\end{subfigure}
	\begin{subfigure}[b]{0.45\linewidth}
		\includegraphics[width=0.95\linewidth]{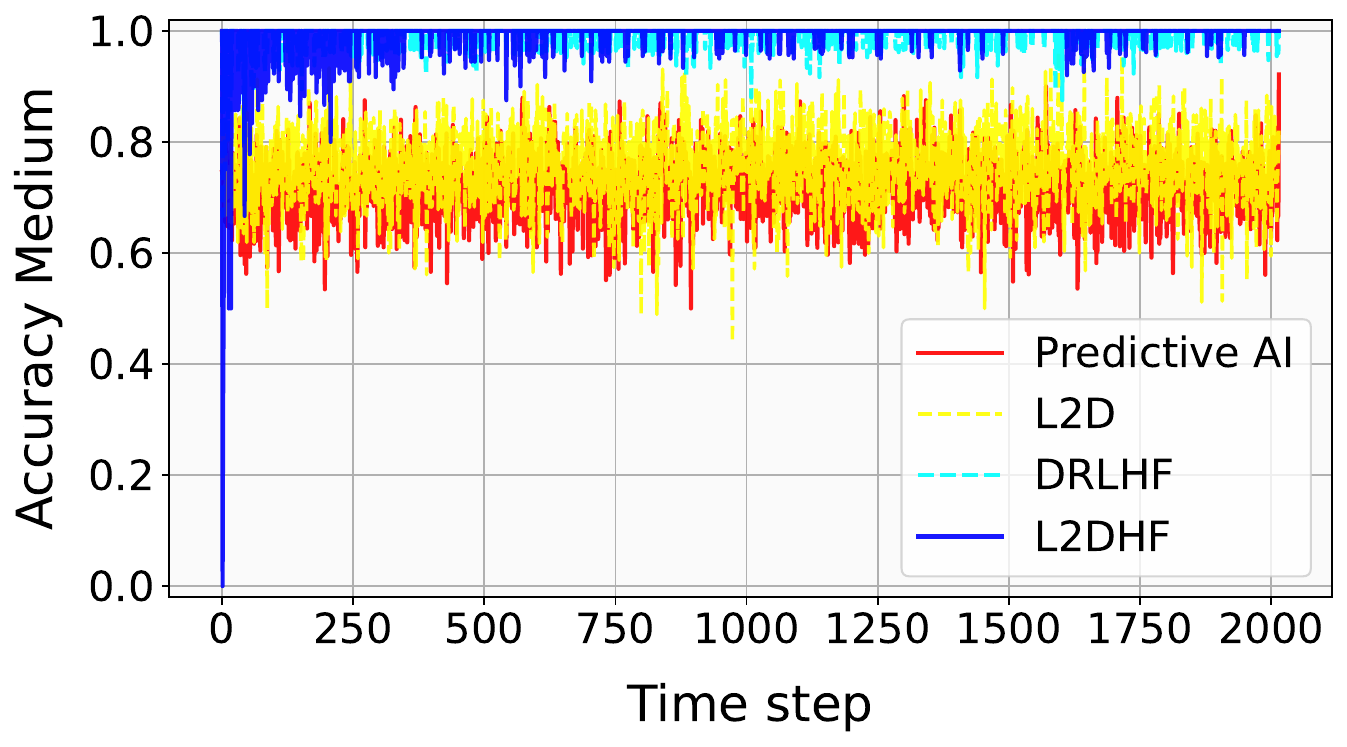}
		\caption{ \footnotesize Accuracy medium category.}\label{fig:Accuracy_Medium_CICIDS}
	\end{subfigure}
	\caption{AP accuracy in categories and overall, CICIDS2017.}
	\label{fig:accuracy_CICIDS}
\end{figure}

Tables \ref{tab:accuracy ensemble+DRHF} and \ref{tab:accuracy ensemble+DRHF_CICIDS} present the overall average AP accuracy as well as the average accuracies across different alert categories, for both processed and unprocessed alerts {by the deferral model}, accounting for the analyst's time budget. Since the predictive AI lacks a deferral model, the total number of alerts it prioritises is reported. Figures \ref{fig:accuracy_UNSW-NB15} and \ref{fig:accuracy_CICIDS} illustrate the models' AP accuracy over time, broken down by individual categories and overall performance. 

Since normal alerts constitute the majority and achieve high accuracy in both datasets, including them can artificially inflate overall accuracy, obscuring the models’ true performance on more severe categories. To address this, we removed normal alerts when calculating the overall accuracy in Figures \ref{fig:accuracy_UNSW-NB15} and \ref{fig:accuracy_CICIDS}. Additionally, Tables \ref{tab:accuracy ensemble+DRHF} and \ref{tab:accuracy ensemble+DRHF_CICIDS} include a column reporting overall accuracy excluding normal alerts. Finally, because the CICIDS2017 dataset does not contain low-category alerts, performance results for this category are unavailable. From these  results, we draw three main insights.

\begin{table}
\centering
\scriptsize

  \caption{Average AP accuracies and alert counts across various alert categories and overall, UNSW-NB15.}
  \label{tab:accuracy ensemble+DRHF}
  \setlength{\tabcolsep}{4pt}
  \begin{adjustbox}{angle=90}
  \begin{tabular}{l c c c c| c c |c c |c c| c c |c c}
    \toprule
    
     \multirow{4}{*}{Model} & & \multicolumn{3}{c}{Overall} &  \multicolumn{2}{c}{Critical} & \multicolumn{2}{c}{High} &  \multicolumn{2}{c}{Medium} & \multicolumn{2}{c}{Low} &  \multicolumn{2}{c}{Normal} \\

    \cmidrule{3-15} 

      & & Count & Accuracy & Accuracy&  Count & Accuracy &  Count & Accuracy &  Count & Accuracy &  Count & Accuracy &  Count & Accuracy\\

      & &  & (with & (without &   &  &   &  &   &  &   &  &   & \\

       & &  & Normal) & Normal) &   &  &   &  &   &  &   &  &   & \\
     
    \midrule
    
\multirow{4}{*}{L2DHF}     & Processed alerts & 792649 & 0.999 & 0.989 &  8607 &0.977 &  11214 &  0.985 &  55483 & 0.992 &  174 & 0.925 &  717171 & 1.0\\

       & (time budget $>0$) & 98.2\%& & & 99.7\%  &  & 98\%  &   & 80\% &  & 75\% &  & 100\%  & \\
     

       & Unprocessed alerts & 14328 & 0.993 & 0.937 & 24  & 0.841 & 196  &  0.918  &  14051 & 0.954  & 57 & 0.390  &  0 & -\\

 & (time budget $=0$) & 1.8\% & &  & 0.3\%  & 
      &  2\% &   & 20\% &  & 25\% &  &  0\% & \\
     

\midrule

\multirow{4}{*}{L2D}     & Processed alerts & 808370 & 0.994 & 0.948 &  8652 &0.864 &  11433 &   0.945 &  69684 & 0.960 &  232 & 0.496 &  718369 & 1.0\\

       & (time budget $>0$) & 100\%& & & 100\%  &  & 100\%  &   & 100\% &  & 100\% &  & 100\%  & \\
     

       & Unprocessed alerts & 0 & 0 & - & 0 & -  & 0  &  -  &  0& -  & 0 & -  &  0 & -\\

 & (time budget $=0$) & 0\% & &  & 0\%  & 
      &  0\% &   & 0\% &  & 0\% &  &  0\% & \\

      \rowcolor{lightgray} P-value& & & -& 0.000 & & 0.000 & & 0.000  & & 0.000 & & 0.000  & & -\\
     

\midrule

      Predictive & Total alerts & 806270 & 0.993 & 0.937 & 8631  & 0.841 & 11410  &  0.918 & 69534   & 0.954 &  231 & 0.390 &  717171 & 1.0\\

      AI & (time budget: \textit{NA}) & & &&   &  &   &   &  &  &  &  &  & \\




     
     \rowcolor{lightgray} P-value& & & -& 0.000 & & 0.000 & & 0.000  & & 0.000 & & 0.000  & & -\\
     
\midrule
\multirow{4}{*}{DRLHF}& Processed alerts & 52238 & 1.0&  0.998 & 803  &0.993 &  700 &  0.997 &  9422 & 0.999 & 8   & 0.875 &  41305 & 1.0\\

& (time budget $>0$) &6.5\% &  & & 9.3\%  &  &  6.1\% &   & 13.6\% &  & 3.4\% &  & 5.8\% & \\


      & Unprocessed alerts & 752378 & 0 & 0 & 7805  & 0 &  10686 &  0 &  59901 & 0 & 224 & 0 & 673762 &  0 \\

     & (time budget $=0$) & 93.5\% & & &  90.7\% &  & 93.9\%  &   & 86.4\% &  &  96.6\% &  &94.2\% & \\
     
\rowcolor{lightgray} P-value& & & - &  0.000 & & 0.000 & & 0.000  & & 0.000 & & 0.641  & & -\\

  \bottomrule
\end{tabular}
\end{adjustbox}
\end{table}


\begin{table*}[h!]
\scriptsize
\centering
  \caption{Average AP accuracies and alert counts across various alert categories and overall, CICIDS2017.}
  \label{tab:accuracy ensemble+DRHF_CICIDS}
  \setlength{\tabcolsep}{4pt}
  \begin{tabular}{l c c c c| c c |c c |c c| c c}
    \toprule
    
     \multirow{4}{*}{Model} & & \multicolumn{3}{c}{Overall} &  \multicolumn{2}{c}{Critical} & \multicolumn{2}{c}{High} &  \multicolumn{2}{c}{Medium} &   \multicolumn{2}{c}{Normal} \\

    \cmidrule{3-13} 

      & & Count & Accuracy & Accuracy&  Count & Accuracy &  Count & Accuracy &  Count & Accuracy &    Count & Accuracy\\

      & &  & (with & (without &   &  &   &  &   &  &   &  \\

       & &  & Normal) & Normal) &   &  &   &  &   &  &   &   \\
     
    \midrule
    
\multirow{4}{*}{L2DHF}     & Processed alerts & 77424 & 0.984 & 0.996 &  348 & 1.0 &  15807 &  1.0 &  40416 & 0.994 &   20853 & 0.952\\

       & (time budget $>0$) & 10\%& & & 63\%  &  &43\%  &   & 33\% &  &  3\%  & \\
     

       & Unprocessed alerts & 731226 & 0.929 & 0.767 & 205  & 0.624 & 20943  &  0.932  &  81407& 0.718  &   628671& 0.968\\

 & (time budget $=0$) & 90\% & &  & 37\%  & 
      &  57\% &   & 67\% &  &  97\% & \\
     

\midrule

\multirow{4}{*}{L2D}     & Processed alerts & 699304 & 0.962 & 0.797 &  450 &0.6 &  28202 &   0.937 &  100203 & 0.758 &   570449 & 0.999\\

       & (time budget $>0$) & 86.9\%& & & 81.4\%  &  & 77.2\%  &   & 82.7\% &  &  88.3\%  & \\
     

       & Unprocessed alerts & 128855 & 0.929  & 0.767 & 103 & 0.624  & 8331  &  0.932   &  21006& 0.718  &   75596 & 0.968 \\

 & (time budget $=0$) & 13.1\% & &  & 18.6\%  & 
      &  22.8\% &   & 17.3\% &  &  11.7\% & \\

      \rowcolor{lightgray} P-value& & & 0.000 & 0.000 & & 0.000 & & 0.000  & & 0.000 &  & 0.000\\
     

\midrule

      Predictive & Total alerts & 806061 & 0.929 & 0.767 & 553  & 0.624 & 36750  &  0.932 & 121823   & 0.718 &   649524 & 0.968\\

      AI & (time budget: \textit{NA}) & & &&   &  &   &   &  &  &   & \\




     
     \rowcolor{lightgray} P-value& & & 0.000 & 0.000 & & 0.000 & & 0.000  & & 0.000 &  & 0.000\\
     
\midrule
\multirow{4}{*}{DRLHF}& Processed alerts & 126191 & 0.994 &  0.993 & 49  &0.959 &  1926 &  0.995 &  57839 & 0.993 &  66377 & 0.995\\

& (time budget $>0$) &15.6\% &  & & 8.9\%  &  &  5.3\% &   & 47.6\% &  & 10.2\% & \\


      & Unprocessed alerts & 680403 & 0 & 0 & 504  & 0 &  34719 &  0 &  63678 & 0 &  581502 &  0 \\

     & (time budget $=0$) & 84.4\% & & &  91.1\% &  & 94.7\%  &   & 52.4\% &  &   89.8\% & \\
     
\rowcolor{lightgray} P-value& & & 0.000 &  0.000 & & 0.000 & & 0.010  & & 0.000 &  & 0.000\\

  \bottomrule
\end{tabular}
\end{table*}

\begin{itemize}
    \item  \textbf{L2DHF consistently outperforms baseline models in AP accuracy across various categories, as well as in the overall performance.} 
    \begin{itemize}[label=$-$]
        \item For \textbf{critical alerts}, the average AP accuracy of the predictive AI and L2D is 0.841 and 0.864 (UNSW-NB15), and 0.624 and 0.6 (CICIDS2017), respectively. L2DHF improves these scores by 16\% and 13\%, raising the accuracy to 0.977 for UNSW-NB15, and by 60\% and 67\%, achieving a perfect 1.0 for CICIDS2017. 

        \item For \textbf{high-category alerts}, the predictive AI and L2D models achieve average AP accuracies of 0.918 and 0.945 (UNSW-NB15), and 0.932 and 0.937 (CICIDS2017), respectively. L2DHF further enhances these scores by 7\% and 4\% for UNSW-NB15, increasing the accuracy to 0.985, and by 7\% for CICIDS2017, achieving a perfect score of 1.0.

        \item For \textbf{medium-category alerts}, the predictive AI and L2D attain accuracies of 0.954 and 0.960 (UNSW-NB15), and 0.718 and 0.758 (CICIDS2017), respectively. L2DHF further improves this by 4\% and 3\%, elevating it to 0.992 for UNSW-NB15, and by 38\% and 31\%, boosting the accuracy to 0.994 (CICIDS2017).
        
        \item Regarding \textbf{low-category alerts}, the predictive AI and L2D achieve accuracies of 0.39 and 0.496 (UNSW-NB15), respectively. L2DHF significantly improves this, reaching 0.925---an increase of 137\% over the predictive AI and 86\% over L2D. 
        \item In terms of \textbf{overall accuracy without normal alerts}, L2DHF reaches 0.989 versus 0.948 for L2D and 0.937 for the predictive AI showing around 4\% and 6\% improvement (UNSW-NB15). L2DHF also achieves 0.996 against 0.797 and 0.767 for L2D and the predictive AI, improving the overall accuracy without normal by almost 25\% and 30\%, respectively (CICIDS2017). 
    \end{itemize}

    \item \textbf{Enhanced predictive AI performance can significantly increase the proportion of alerts processed by L2DHF, thereby boosting the percentage of alerts with improved AP accuracy}. As noted earlier, not all alerts may be processed by L2DHF due to the analyst’s time constraints. In the UNSW-NB15 dataset, the predictive AI achieves a perfect accuracy of 1.0 for normal alerts, meaning these alerts require no further refinement, and are excluded from being forwarded to the DRLHF component of L2DHF. Consequently, only severe alerts are sent to L2DHF, resulting in a significantly higher percentage of processed alerts for UNSW-NB15. Specifically, 99.7\% of critical alerts are processed by L2DHF for UNSW-NB15, compared to 63\% for CICIDS2017. The processed critical alerts achieve high accuracies of 0.977 (UNSW-NB15) and 1.0 (CICIDS2017), while the predictive AI’s average accuracy, which is 0.841 (UNSW-NB15) and 0.624 (CICIDS2017), is applied for the unprocessed critical alerts. This highlights the AP accuracy improvement for processed alerts.

    
    As shown in Table \ref{tab:accuracy ensemble+DRHF_CICIDS}, in CICIDS2027, L2DHF achieves a lower accuracy for normal alerts (0.952) compared to the predictive AI (0.968). However, this figure is misleading, as only 3\% of normal alerts were processed by L2DHF, making its accuracy value unreliable. To address this, we could exclude normal alerts from being sent to DRLHF part of L2DHF for CICIDS2017 as well. While the predictive AI’s accuracy for normal alerts is not perfect for CICIDS2017, it is reasonably high. This exclusion would increase the number of processed severe alerts, which are more important, thereby leading to a significant rise in the total number of processed alerts and improved AP accuracy. 

    The trend of more processed alerts with higher predictive AI accuracy is also observed for L2D. Although normal alerts are not excluded from L2D processing in UNSW-NB15, the predictive AI's higher AP accuracy enables L2D to process 100\% of alerts, whereas this value drops to 86.9\% for CICIDS2017. This is because higher AP accuracy from the predictive AI increases L2D's confidence in its deferral decisions, allowing more alerts to be accepted with the predictive AI’s assigned priority. As a result, fewer alerts are deferred to the analyst, optimising analyst time for more uncertain cases and boosting the number of processed alerts.

    \item \textbf{DRLHF processes significantly fewer alerts compared to the other models, resulting in a large volume of un-prioritised alerts.} As shown in Table \ref{tab:accuracy ensemble+DRHF}, for UNSW-NB15, DRLHF prioritises only 6.5\% of total alerts, leaving the remaining 93.5\% unprocessed and therefore un-prioritised, as DRLHF lacks initial prioritisation from the predictive AI. This pattern is also evident across different alert categories. For instance, only 9.3\% of critical alerts, 6.1\% of high-category alerts, and 3.4\% of low-category alerts are processed by DRLHF. In the case of CICIDS2017, as shown in Table \ref{tab:accuracy ensemble+DRHF_CICIDS}, although the total number of processed alerts increases to 15.6\%, the majority of top-priority alerts, such as critical and high-category alerts, which pose greater vulnerabilities and are therefore more important for SOCs, remain largely unprocessed and without assigned priority. Only 8.9\% of critical alerts and 5.3\% of high-category alerts are processed. 
    
    Table \ref{tab:accuracy ensemble+DRHF} shows that DRLHF slightly outperforms L2DHF in AP accuracy for some severe categories. However, this higher accuracy applies to a much smaller portion of the alerts. For instance, in Table \ref{tab:accuracy ensemble+DRHF}, while DRLHF achieves an accuracy of 0.997 for high-category alerts, it only covers 6.1\% of them, leaving 93.9\% unprocessed and without assigned priority. In contrast, L2DHF achieves an accuracy of 0.985 across 98\% of high-category alerts. For the remaining 2\% of high-category alerts not processed by L2DHF, the predictive AI's accuracy (0.918) is applied. Therefore, while DRLHF may show seemingly higher accuracy in some cases, this is misleading, as it processes a far smaller proportion of the alerts and cannot be considered superior to L2DHF.

    L2D processes higher percentage of alerts. However, due to its suboptimal accuracy, especially for critical, high, medium and low categories, these alerts still require more careful analyst review, increasing the analyst's workload.

\end{itemize}

Minor variations in the total number of alerts submitted to the models arise from the random Poisson-distributed alert arrivals at each time steps. To assess the significance of accuracy differences between L2DHF and baselines, we performed the Mann-Whitney U-test~\citep{MacFarland2016}. The results, presented in Tables \ref{tab:accuracy ensemble+DRHF} and \ref{tab:accuracy ensemble+DRHF_CICIDS}, show that all P-values, except for one case (low-category alerts for DRLHF in UNSW-NB15), are under 0.05, indicating statistical significance at the 95\% confidence level. 
Since all models achieve perfect accuracy for normal alerts in UNSW-NB15, comparisons of overall accuracy with normal alerts were omitted.

\subsection{Misprioritisations}
\begin{figure}[t!]
  \centering
  \includegraphics[width=0.85\linewidth]{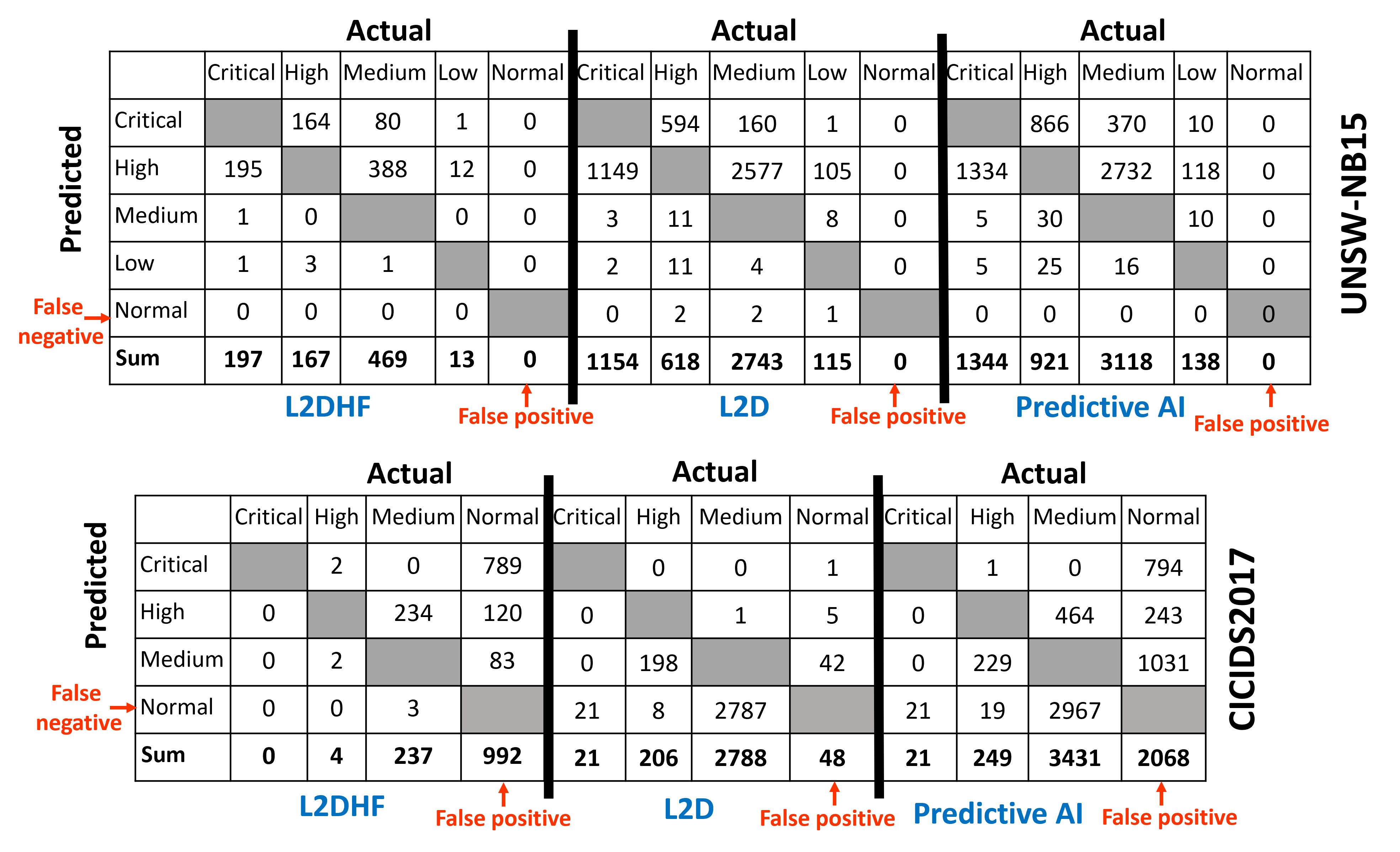}
  \caption{The count of  misprioritisations.}
  \label{false AP}
\end{figure}

Figure  \ref{false AP} shows the count of misprioritisations over time steps across all categories. DRLHF is excluded from the comparison due to its limited number of processed alerts in most categories, making it impractical to compare with other models. For a fair comparison, we ensure that the processed alert percentage of L2DHF matches that of the other models. 

\begin{itemize}[label=$-$]
    \item In general, L2DHF notably reduces misprioritisations,  including misprioritisation between severe threat levels, false positives and false negatives, across all categories compared to other models, with the exception of L2D on CICIDS2017, which demonstrates a lower false positive number. 

    \item In the case of \textbf{critical alerts} that are most important for SOCs to correctly detect, L2D and the predictive AI misprioritise 1,154 and 1,344 alerts, respectively, in UNSW-NB15. In contrast, L2DHF significantly reduces misprioritisations of critical alerts to 197, achieving reductions of 83\% and 85\% compared to L2D and predictive AI, respectively. In CICIDS2017, L2DHF achieves a 100\% reduction in misprioritisation of critical alerts, compared to 21 misprioritised critical alerts by both L2D and the predictive AI. This improvement is particularly important, as misprioritised critical alerts can lead to serious breaches or delayed responses to active threats.

    \item In a similar pattern for \textbf{high-category alerts}, L2DHF misprioritises 167 alerts in UNSW-NB15, reflecting a reduction of nearly 73\% and 82\%, compared to the 618 misprioritised by L2D and the 921 misprioritised by the predictive AI, respectively. In CICIDS2017, L2DHF misprioritises only 4 high-category alerts, compared to 206 and 249 misprioritised by L2D and the predictive AI, respectively, resulting in a 98\% reduction in misprioritisation of high-category alerts. 

    \item \textbf{L2DHF not only reduces misprioritisations, but also mitigates the danger of more severe alerts being misclassified into less severe categories, thereby reducing the likelihood of top-severity alerts being overlooked.} For example, in UNSW-NB15, L2DHF primarily misclassifies critical alerts as high-priority alerts, which still receive significant attention from SOCs for investigation and response. Only 1 critical alert is misclassified as medium and another 1 as low by L2DHF. In contrast, the predictive AI misclassifies 5 critical alerts as medium priority and another 5 as low priority. In CICIDS2017, all 21 critical alerts misprioritised by L2D and the predictive AI are labelled as normal priority, substantially increasing the potential harmful consequences of overlooking these critical-severity threats. Moreover, in CICIDS2017, among the misprioritised high-category alerts, 2 are downgraded to medium by L2DHF, while L2D downgrades 198 to medium and 8 to normal, and the predictive AI downgrades 229 to medium and 19 to normal. Such misprioritisations can introduce serious security incidents, especially if a critical or high-category alert is incorrectly treated as low priority and left unaddressed.

    \item In terms of \textbf{false positives}, in UNSW-NB15, all three models: L2DHF, L2D, and the predictive AI produce zero false positives, likely due to the predictive AI’s perfect accuracy in initially prioritising normal-category alerts.  

    In CICIDS2017, L2DHF generates 992 false positives, representing a 52\% reduction compared to the 2,068 false positives produced by the predictive AI. In contrast, L2D produces only 48 false positives, which may seem favourable at first glance. However, this lower count is misleading, as L2D results in a substantially higher number of misprioritised severe alerts than L2DHF. For instance, 2,788 misprioritisations compared to 237 in the medium category, and 206 misprioritisations compared to 4 in the high category. Given the analyst's limited time, models can only defer a subset of alerts for human correction. L2D defers more normal-category alerts to the analyst, leaving many severe alerts unreviewed. This leads to fewer false positives but a significantly higher number of  misprioritisations for severe alerts. In contrast, L2DHF focuses on deferring more severe alerts to the analyst, helping reduce misprioritisations in these crucial categories. Although this results in more normal alerts going unreviewed, leading to a higher false positive count, L2DHF's result is preferable in terms of maintaining overall system security and operational effectiveness. 

    \item In terms of \textbf{false negatives}, in UNSW-NB15, both L2DHF and the predictive AI result in zero false negatives, while L2D has only a negligible number, probably because the predictive AI consistently prioritises normal-category alerts with complete accuracy. In CICIDS2017, L2DHF significantly outperforms both L2D and the predictive AI. While L2D and the predictive AI result in 2,816 (21 critical + 8 high + 2,787 medium) and 3,007 (21 critical + 19 high + 2,967 medium) false negatives respectively, L2DHF reduces this number to just 3, representing an almost 100\% reduction of false negatives.

\end{itemize}

\subsection{Unprocessed alerts and deferred alerts by the deferral model}
This section examines the number of unprocessed alerts by the deferral model and the number of deferred alerts sent to the analyst, offering a clearer understanding of model performance in enhancing AP and reducing analyst workload. Our previous analyses have shown that alerts not processed by the DRL agent (acting as the deferral model) tend to have significantly lower AP accuracy. This can substantially increase the analyst’s workload during subsequent investigation steps, as more detailed analysis is needed to identify any missed critical alerts. Therefore, models that enable the DRL agent as the deferral model to process a higher number of alerts generally demonstrate better overall performance.

\begin{figure}[ht!]
  \centering
  \includegraphics[width=0.8\linewidth]{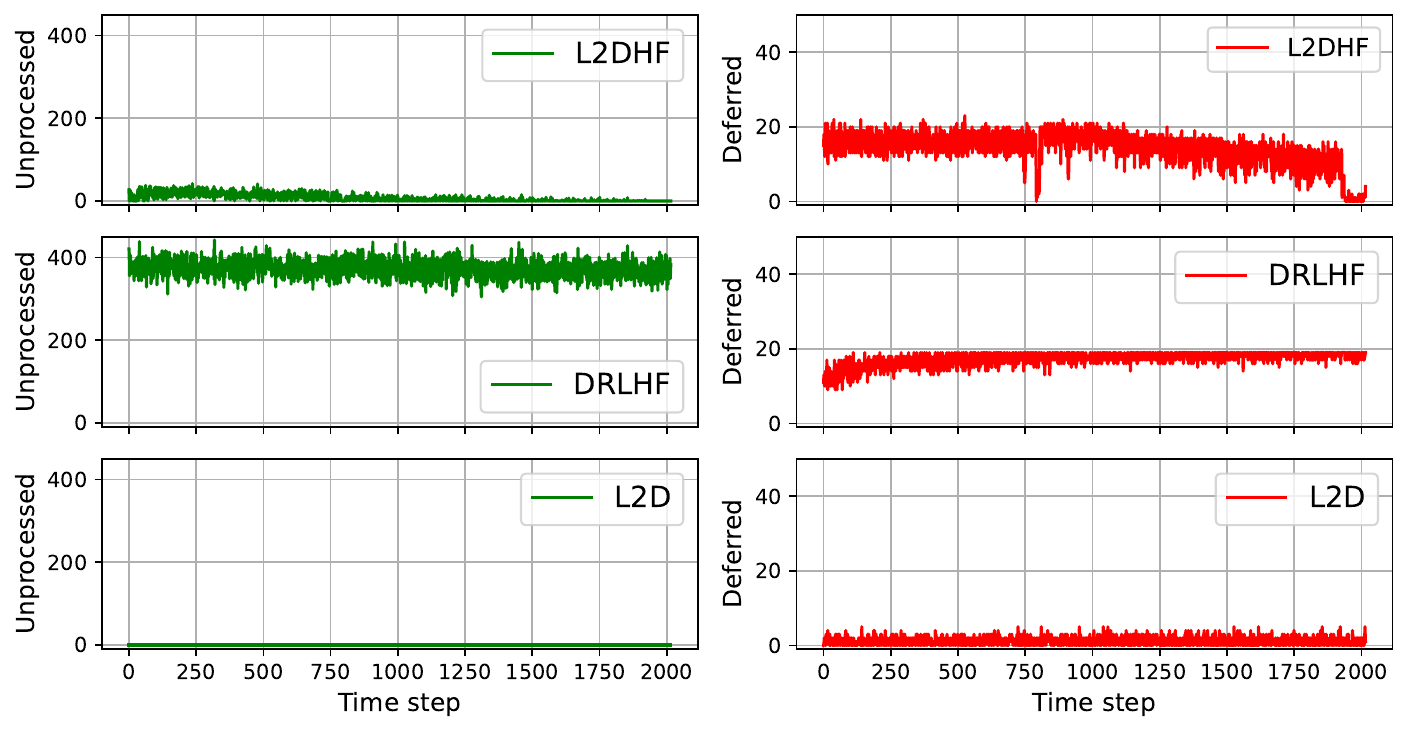}
  \caption{Number of unprocessed alerts by the deferral models (left) and deferred alerts to the analyst (right) over time, UNSW-NB15.}
  \label{fig:unaddressed alerts}
\end{figure}

\begin{figure}[t!]
  \centering
  \includegraphics[width=0.8\linewidth]{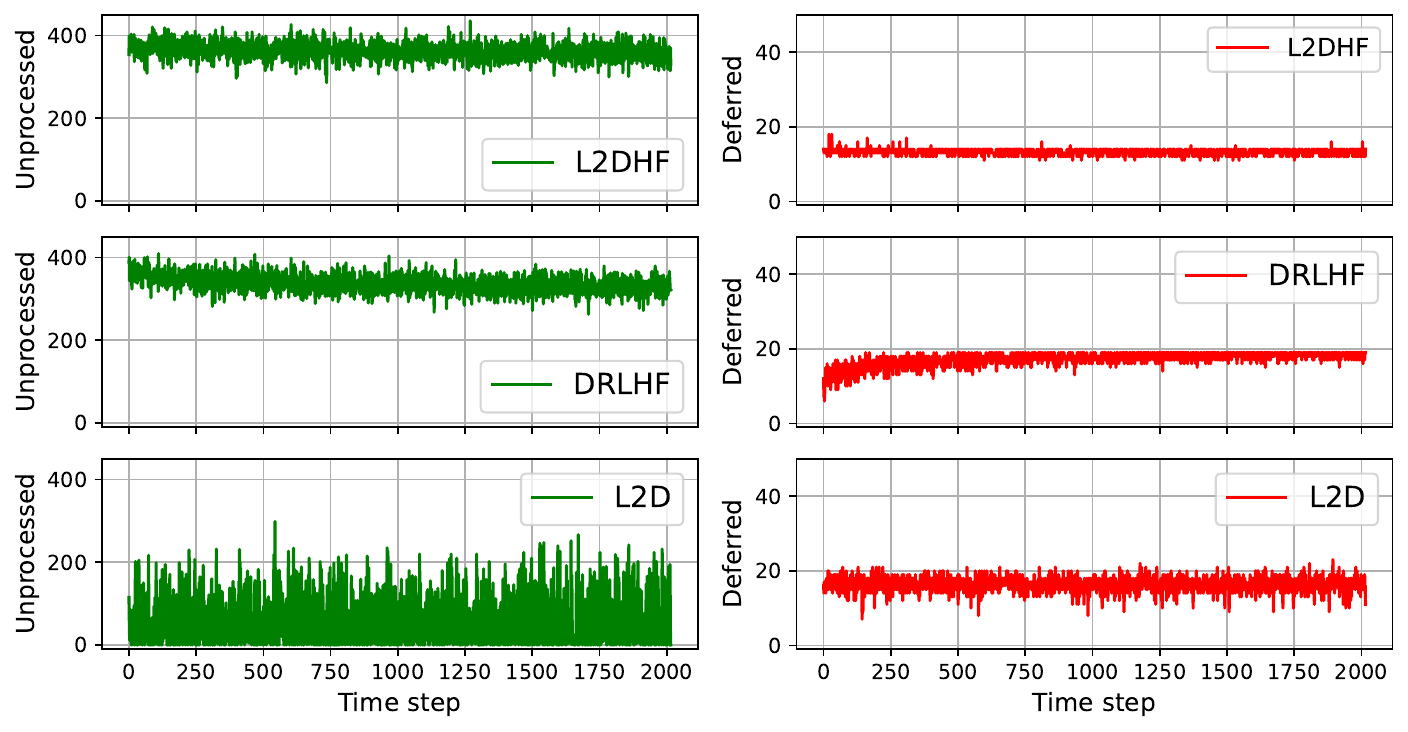}
  \caption{Number of unprocessed alerts by the deferral models (left) and deferred alerts to the analyst (right) over time, CICIDS2017.}
  \label{fig:unaddressed alerts, CICIDS}
\end{figure}

Figures \ref{fig:unaddressed alerts} and \ref{fig:unaddressed alerts, CICIDS} show the number of unprocessed  alerts by the deferral model, along with the number of alerts deferred to the analyst over time for UNSW-NB15 and CICIDS2017, respectively\footnote{The predictive AI is not included in this analysis as it lacks a deferral model.}. Figures \ref{fig:unprocessed and deferred, boxplot, unsw} and \ref{fig:unprocessed and deferred, boxplot, CICIDS} provide comparative boxplots of unprocessed and deferred alerts across models. Key insights are discussed below:

\begin{itemize}[label =$-$]
    \item \textbf{L2DHF Performance:} L2DHF eliminates unprocessed alerts for UNSW-NB15 by 100\% over time, dropping from an average of 17 in the first 500 time steps to 0 in the last 500. However, for CICIDS2017, unprocessed alerts remain consistently high, averaging around 363, with no noticeable decline due to the large volume of alerts sent to the DRL agent, which exceeds the number that can be deferred to the analyst because of the analyst limited time. This issue could be mitigated by incorporating multiple analysts into L2DHF, as will be discussed later. 

    In UNSW-NB15, L2DHF decreases deferred alerts from an average of 16 in the first 500 time steps to 10 in the last 500, marking a 37\% reduction over time and significantly lowering the analyst's workload. Figure \ref{fig:unprocessed and deferred, boxplot, unsw} shows that L2DHF consistently results in fewer deferred alerts versus DRLHF, with an average of 14 compared to 17 in DRLHF, representing an approximate 18\% reduction in average analyst workload compared to DRLHF. In CICIDS2017, the number of deferred alerts remains constant averagely at 13 across both the first and last 500 time steps. Although no reduction is observed over time in this case, the value of 13 is the minimum average overall among the evaluated methods, resulting in a 19–23\% reduction in analyst workload when compared to DRLHF (average deferred alerts: 17) and L2D (average deferred alerts: 16), as shown in Figure \ref{fig:unprocessed and deferred, boxplot, CICIDS}.

    \item \textbf{DRLHF Performance:} Similar patterns emerge across both datasets with DRLHF. The number of unprocessed alerts remains high, averaging around 373 for UNSW-NB15 and 338 for CICIDS2017. Additionally, DRLHF increases the number of deferrals over time to maximise the utilisation of the analyst’s time budget to improve AP, leading to a rise in deferred alerts over time. Deferred alerts increase from an average of 15 for UNSW-NB15 and 16 for CICIDS2017 in the first 500 time steps to 18 in the last 500, marking a 20\% and 12.5\% rise over time, respectively, and subsequently increasing the analyst's workload. Figures \ref{fig:unprocessed and deferred, boxplot, unsw} and \ref{fig:unprocessed and deferred, boxplot, CICIDS} also show that DRLHF results in the highest average number of deferred alerts among all models, thereby imposing the highest workload.

    \item \textbf{L2D Performance:} In L2D, the average number of deferred alerts remains constant over time for both datasets, as shown in Figures \ref{fig:unaddressed alerts} and \ref{fig:unaddressed alerts, CICIDS}. For CICIDS2017, a larger number of alerts have predictive AI priorities below the confidence threshold, prompting the model to maximise deferrals in an attempt to fully utilise the analyst's time for priority revision. As a result, the average number of deferred alerts reaches 16. Additionally, unprocessed alerts average 52. In contrast, for UNSW-NB15, the number of deferred and unprocessed alerts in L2D is minimal compared to L2DHF as illustrated in Figure \ref{fig:unprocessed and deferred, boxplot, unsw}. However, considering L2D’s relatively poor AP accuracy compared to L2DHF, this indicates inefficiencies in how analyst time is utilised to improve AP. Deferring fewer alerts does not necessarily indicate better model efficiency.

\end{itemize}

\begin{figure}[t!]
  \centering
  \includegraphics[width=0.9\linewidth]{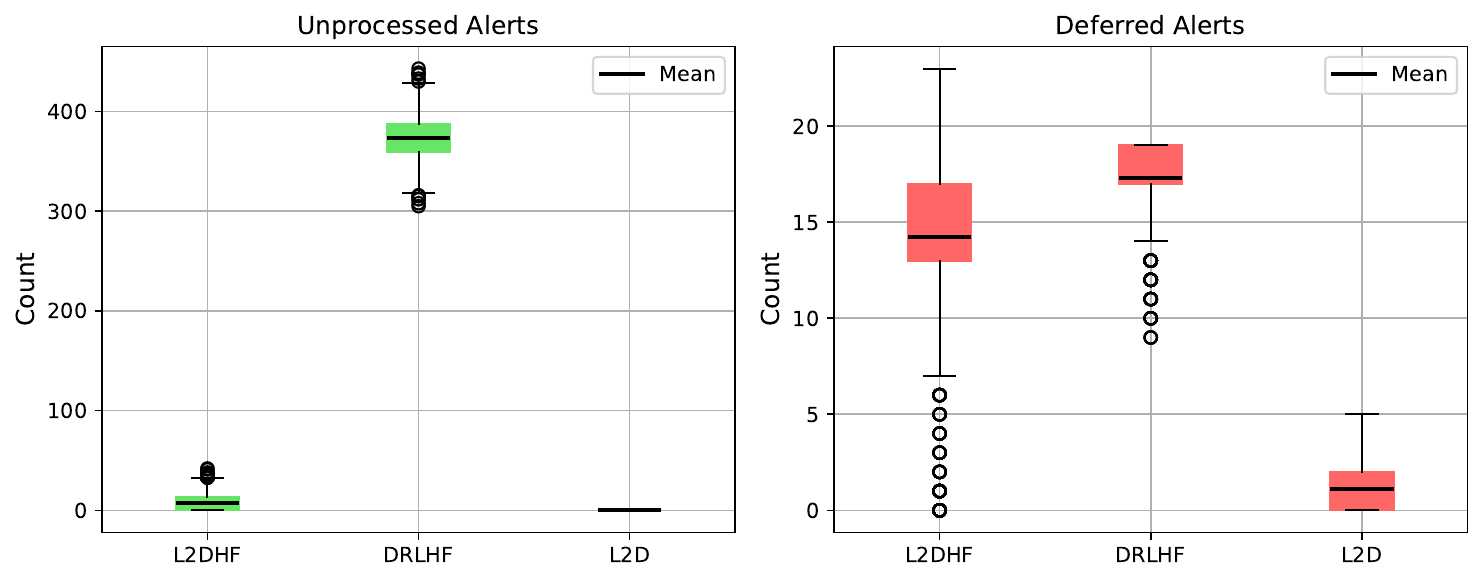}
  \caption{Comparative boxplots of  unprocessed alerts (left) and  deferred alerts (right), UNSW-NB15.}
  \label{fig:unprocessed and deferred, boxplot, unsw}
\end{figure}

\begin{figure}[t!]
  \centering
  \includegraphics[width=0.9\linewidth]{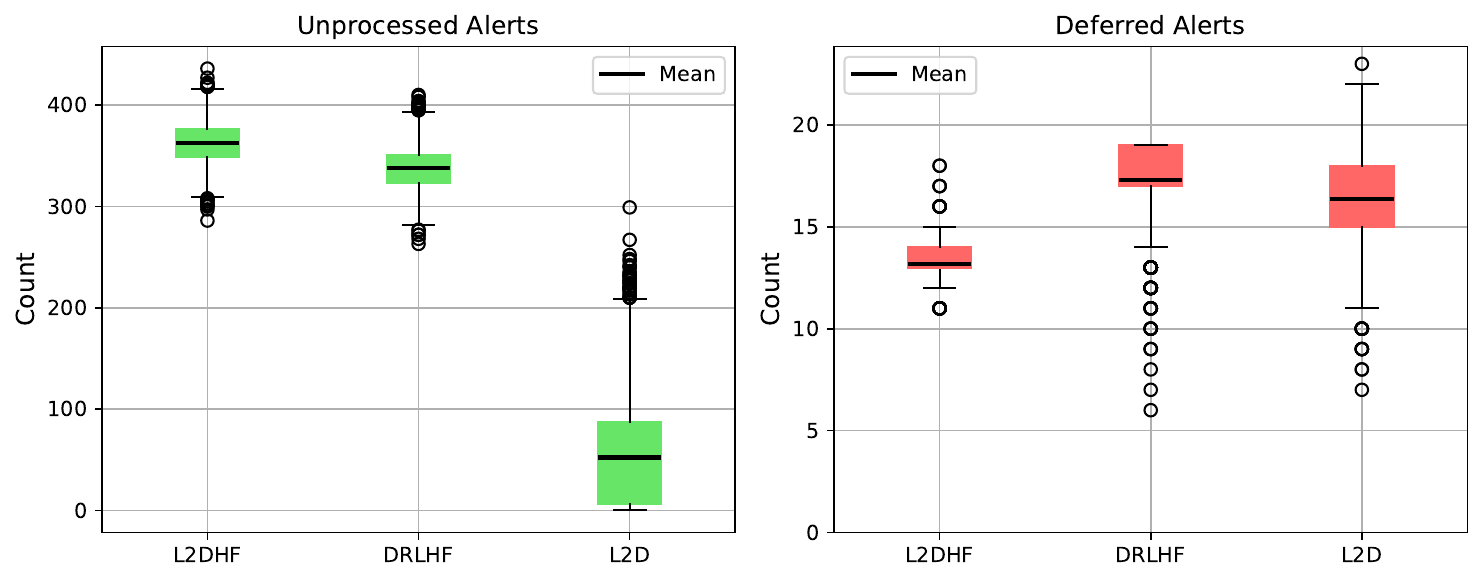}
  \caption{Comparative boxplots of  unprocessed alerts (left) and  deferred alerts (right), CICIDS2017.}
  \label{fig:unprocessed and deferred, boxplot, CICIDS}
\end{figure}

Based on these findings, we
conclude that L2DHF is the most effective model. It enhances AP performance by decreasing the number of unprocessed alerts, while simultaneously reducing deferred alerts, thus alleviating analyst workload.

The challenge of unprocessed alerts stems from the analyst’s limited time budget. Given more time, the analysts could process all alerts. One possible solution is to increase the number of analysts, allowing the deferral model to distribute alerts more effectively. However, this solution comes with the trade-off of higher personnel costs, as more analysts would incur higher labour costs. This study models the scenario with a single analyst. Due to its superior AP accuracy and the reduction in deferred alerts, which minimises the
need for additional analysts, L2DHF emerges as the preferred model for handling unprocessed alerts, outperforming both DRLHF and L2D.\\

\noindent \textbf{Note:} It is important to emphasise that L2DHF is not intended to replace human judgement, but to complement it. Human analysts remain indispensable due to the complex and evolving nature of SOC operations. L2DHF aims to alleviate their workload while maximising AP performance. By deferring uncertain and novel cases to human experts, L2DHF helps minimise errors, learns from human feedback to optimise its deferral strategy, and improves AP accuracy. L2DHF also mitigates analyst workload by optimising the number of deferred alerts.

\subsection{Execution time}
\begin{figure}[t!]
	\centering
	\begin{subfigure}[b]{0.45\linewidth}
		\includegraphics[width=1\linewidth]{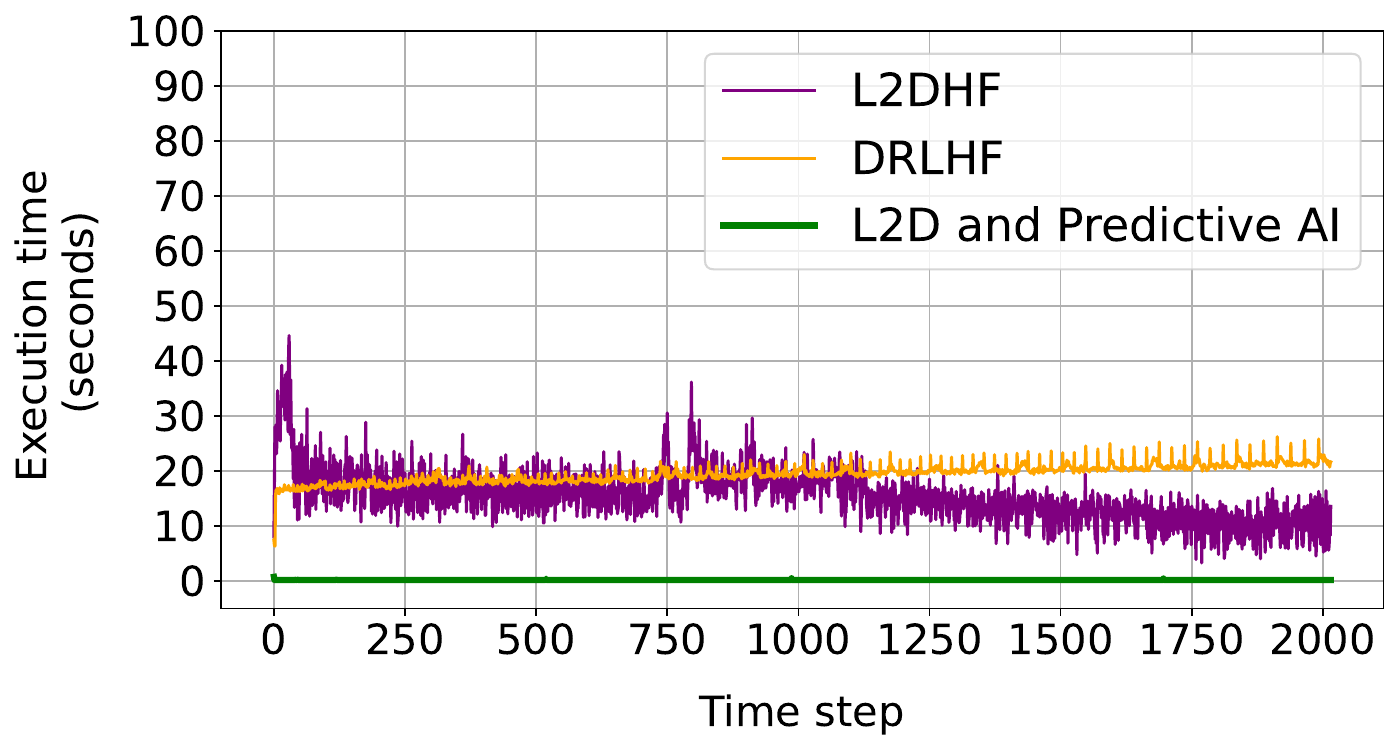}
		\caption{\footnotesize   UNSW-NB15. }\label{fig:ExeTime_UNSW-NB15}
	\end{subfigure}
	\begin{subfigure}[b]{.45\linewidth}
		\includegraphics[width=1\linewidth]{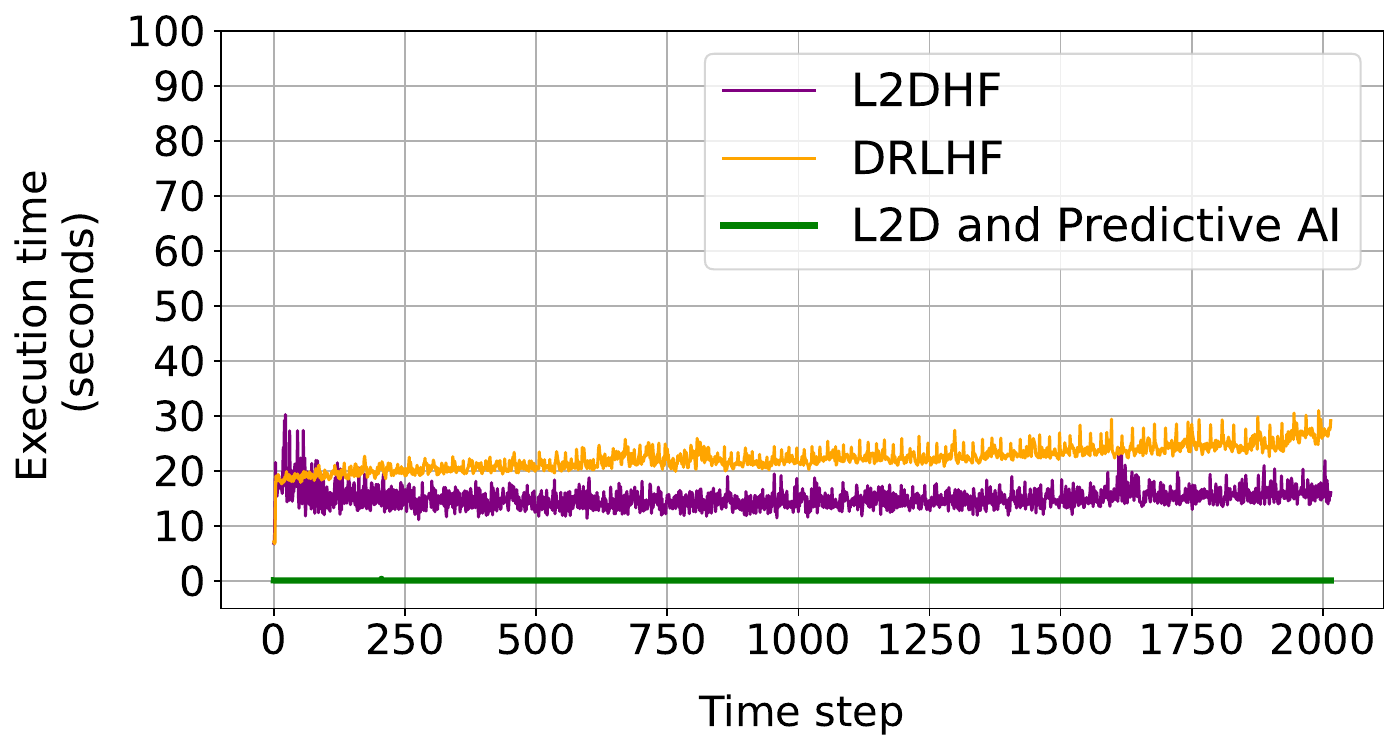}
		\caption{ \footnotesize CICIDS2017.}\label{fig:ExeTime_CICIDS}
	\end{subfigure}
	\caption{Execution time of models.}
	\label{fig:execution_time}
\end{figure}

Figure \ref{fig:execution_time} illustrates the execution times of the models across time steps for both datasets, reflecting the time required to handle incoming alerts at each step. L2DHF consistently maintains efficient execution times, ranging from approximately 10 to 40 seconds across time steps for both UNSW-NB15 and CICIDS2017. These times represent the total execution time of each model’s AI module. For instance, L2DHF’s AI module includes both the predictive AI and the DRL agent, while DRLHF’s consists solely of the DRL agent. The remaining duration within each time step is available for the analyst to review alerts. This demonstrates L2DHF's suitability for real-time AP tasks, aligning well with the speed of alert generation in SOCs and the urgency of their prioritisation. DRLHF shows an increasing execution time over the time steps, with higher execution times than those of L2DHF after time step 1000 for UNSW-NB15 and in most time steps for CICIDS2017.

\section{Threats to validity} \label{sec: limit}

This section outlines the key threats to the validity of our study, particularly focussing on the challenges of involving real human analysts, employing realistic datasets in the experimental setup, and assumptions related to analyst feedback and AVAR reliability.

\subsection{Human analyst involvement} \label{sec:human} 
Integrating real human feedback poses a major challenge for RLHF frameworks. As noted by \cite{christiano2017deep}, providing real human feedback as a direct reward is often impractical, given RLHF systems typically require hundreds or thousands of interactions to learn effectively. Reducing the number of interactions is necessary to make the training feasible with real human feedback but may come at the cost of learning efficiency and overall performance of RLHF models. This issue is further amplified in the SOC context, where security analysts are highly skilled professionals whose time is both limited and valuable. Engaging security analysts in prolonged interactive experiments over extended periods, such as weeks, would be not only impractical but also disruptive to their operational duties. 

To address this, we followed an established and pragmatic alternative of using ground truth labels as a proxy for analyst feedback \citep{mozannar2020consistent,cao2024reinforcement}. This approach has also been adopted in a SOC-related research \citep{wang2024combating}, where ground truth was used to emulate analyst input in an RLHF-enabled anomaly detection system. Although it may not fully capture the subjectivity of human decision-making, this practice facilitates experimentation at scale while mitigating the cost and complexity of involving experts.

\subsection{Realistic datasets} \label{sec:dataset2}
Effective experimentation for AP in SOCs necessitates access to cybersecurity datasets that reflect realistic operational conditions. Ideally, it would involve real-world SOC data; however, realistic datasets are rarely available due to confidentiality and security constraints. Acquiring realistic data is challenging due to the need for diverse, up-to-date attacks, realistic operational settings, and traffic characteristics reflecting the real-world conditions (with errors, imbalanced, etc.)  \citep{duraz2023cyber}. 

The datasets used in this study, UNSW-NB15 and CICIDS2017, have been widely adopted in ML-based SOC and cybersecurity research \citep{wu2025boosting,kumar2025nids,binbusayyis2024hybrid}. They include environments and traffic patterns that are representative of real-world operational conditions \citep{duraz2023cyber}. 

\subsection{Analyst error and AVAR reliability} \label{sec:analyst-AVAR limitations}
Our study assumes that human analysts always provide correct feedback when reviewing alerts. While this assumption simplifies the experimental design and facilitates focused evaluation, it does not reflect the variability in analyst decisions that exists in real SOCs, where even expert analysts are susceptible to error, especially under conditions of cognitive load. Moreover, any incorrect analyst decisions are stored in AVAR without opportunities for revision. This could degrade the long-term reliability of AVAR. 


Future work could address these limitations by explicitly modelling the potential for analyst error. Additionally, mechanisms to ensure the quality of AVAR data could be implemented. For example, alerts could be reviewed by multiple analysts before being permanently recorded, or regular audits by domain experts could be introduced to filter inaccurate entries and maintain the integrity of AVAR.
\section{Conclusion} \label{sec:conc}
This paper introduced \textit{Learning to Defer with Human Feedback }(L2DHF), a novel approach to human-AI teaming (HAT) aimed at improving alert prioritisation (AP) in security operation centres (SOCs). Central to L2DHF is an adaptive deferral model powered by DRLHF, enabling the system to continuously refine its deferral decisions based on human feedback. This leads to improved AP accuracy, reduced misprioritisations, and decreased analyst workload. 

Experimental results on the UNSW-NB15 and CICIDS2017 datasets show consistent improvements in AP performance. L2DHF enhanced AP accuracy for critical alerts by 13-16\% on UNSW-NB15 and 60-67\% on CICIDS2017 compared to baseline models. On the CICIDS2017 dataset, L2DHF achieved a 100\% reduction in misprioritisaions of critical alerts, a 98\% reduction in misprioritisaions of high-category alerts, and a 52\% reduction in false positives versus the baselines. Additionally, L2DHF decreased the number of deferred alerts by 37\% over time on UNSW-NB15, helping reduce analyst workload. Its execution time remained highly efficient, supporting its suitability for real-time AP in SOCs.

Future work could address the challenge of unprocessed alerts by leveraging multiple analysts, with mechanisms for assigning alerts based on expertise and preferences. Further research might explore how the DRL agent could provide personalised feedback to analysts, fostering a reciprocal loop to improve both deferrals and analyst performance. Another promising direction is the integration of large language models (LLMs) to support HAT in AP. LLMs could enable natural language interactions between analysts and AI systems, generate alert summaries, and provide interpretable explanations to assist in decision-making.

\section*{Acknowledgements}
This work was supported by CSIRO’s Collaborative Intelligence (CINTEL) Future Science Platform.

\section*{Declaration of AI use}

The authors confirm that AI and AI-assisted technologies were used during the preparation of this work. Specifically, OpenAI’s ChatGPT was used to assist with language editing, paraphrasing, and improving readability in certain parts of the manuscript. The authors have reviewed and taken full responsibility for the content.

\small
\setlength{\bibsep}{0.0pt}
\typeout{}\bibliography{L2DHF}

\bibliographystyle{model5-names}\biboptions{authoryear}

\end{document}